\documentclass[a4paper,10pt]{article}

\usepackage{amssymb}
\usepackage{amsmath}
\usepackage[usenames,dvipsnames]{color}
\usepackage{hyperref}
\usepackage[left=.8in,right=.8in,top=.8in,bottom=.8in]{geometry}              

\usepackage{xcolor}
\usepackage{mathpazo}
\usepackage[normalem]{ulem}
\usepackage{palatino}
\usepackage{mathrsfs}
\usepackage[mathscr]{euscript}
\usepackage{lscape}
\usepackage{cancel}
\usepackage{bm}
\usepackage{bbm}
\usepackage{multirow} 
\usepackage{amsfonts}
\usepackage{amssymb}
\usepackage{graphicx}
\usepackage{amsmath}
\usepackage[usenames,dvipsnames]{color}
\usepackage{hyperref}
\usepackage{epstopdf,epsfig}
\usepackage{color}

\usepackage{cite}

\hypersetup{
    colorlinks=true,         % false: boxed links; true: colored links, false is default
    linkcolor=MidnightBlue,          % color of internal links, red is default
    citecolor=BrickRed,        % color of links to bibliography, 'green' is default
    urlcolor=MidnightBlue             % color of external links, cyan is default
}

\newtheorem{defi}{Definition}

\newtheorem{cor}{Corollary}

\newcommand{\st}[1]{\textrm{\tiny{#1}}}

\newcommand{\diby}[2]{\ensuremath{\frac{\delta #1}{\delta #2}}}

\newcommand{\order}[1]{\ensuremath{\mathcal{O}(#1)}}

\newcommand{\rd}{{\mathrm{d}}}

\def\be{\begin{equation}}
\def\ee{\end{equation}}
\def\bea{\begin{eqnarray}}
\def\eea{\end{eqnarray}}

\def\sfrac{\textstyle \frac }
\title{Local(ish) gravity theories in  conformal superspace.}
\author{\bf Henrique Gomes\footnote{\href{mailto:gomes.ha@gmail.com}{gomes.ha@gmail.com}}\\\it Perimeter Institute for Theoretical Physics\\ \it 31 Caroline Street, ON, N2L 2Y5, Canada}
\begin{document}
\maketitle
\begin{abstract}
The conformal method for the initial value formulation of GR is one of the most powerful and widely used tools in both analytical studies and numerical simulations of solutions. As is well-known, it exploits a hidden spatial conformal symmetry that can be extracted from the equations. But why conformal symmetry?  Motivated by these questions and by well-known obstacles to quantum gravity,  I look for the most general geometrodynamical symmetries compatible with a reduced physical configuration space for metric gravity.  I argue that they are indeed spatial conformal diffeomorphisms.  The next question is: what sort of theories most naturally incorporate this symmetry?  Demanding locality for an action for metric gravity compatible with these principles determines the allowed operators, both for the purely gravitational part as well as matter couplings. The symmetries guarantee that there are two gravitational propagating physical degrees of freedom, but no explicit refoliation-invariance. The simplest such system has a geometric interpretation as a geodesic model in infinite-dimensional conformal superspace. One example of solution to the equations of motion corresponds to a static Bianchi IX spatial ansatz. The unique coupling to electromagnetism forces the electromagnetic  equations to be hyperbolic, enabling us to \lq\lq{}build\rq\rq{}  a standard space-time causal structure. There are, however, deviations from the standard Maxwell equations when space-time anisotropies become too large. 
 Regarding quantization, with the geometric interpretation and the lack of refoliation invariance, the path integral treatment of the symmetries becomes much less involved than the similar approaches to GR. The symmetries form an (infinite-dimensional) Lie algebra, and no BFV treatment is necessary. Moreover, one can use gauge-invariant variational principles \textit{for selecting the boundary conditions} of the path integral. We find that the propagator around the homogeneous solution has up to 6-th order spatial derivatives, giving it plausible regularization properties as in Horava-Lifschitz.   
\end{abstract}

%\tableofcontents

\section{Introduction}
\subsection{The York method and conformal symmetry.}

   When we want to make predictions about the future behavior of gravitational degrees of freedom, we require an expression of the Einstein equations in terms of evolution equations, i.e. making explicit reference to a time function. Roughly, given an initial value in some space-like surface, these equations should tell us what to expect on some later such surface. This obligates us to break up the original spacetime covariant picture into  a foliation of space-time into spatial hypersurfaces (see fig. 1). We don't need to reinvent this account, as it is already standard in the study of general relativity, going by the acronym ADM  (after Arnowitt, Deser and Misner \cite{ADM}, see also  \cite{3+1_book} for a modern review of the methods).   Indeed most of the work in numerical GR requires the use of the dynamical approach.  Correspondingly, assuming the spacetime manifold is of the form $\Sigma\simeq M\times \mathbb R$, we perform a 3+1 split of the spacetime metric:
\be\label{3+1_eq}
{}^\st{(4)}g_{\mu\nu} = \left(
\begin{array}{cc} - {N^2} +  {g_{ab}}\, {\xi^a \, \xi^b} &   {g_{ak}} \, {\xi^k}\\   {g_{jk}} \, {\xi^k} &   { g_{ab}}
\end{array} \right) \,,
\ee
where we used Greek letters to denote spacetime tensor components, and Latin letters for the spatial ones. The Einstein-Hilbert action is decomposed according to \eqref{3+1_eq},
$$
\displaystyle \int d^4 x \sqrt { {}^\st{(4)}g}  ~ {}^\st{(4)}R  = \int dt d^3 x \left( \dot g_{ab} p^{ab} +{N} \, \mathcal H[g,p] + {\xi^a} \mathcal D_a[g,p]  \right).
$$
The lapse $N$ and shift $\xi^a$ have no conjugate momenta, and are thus Lagrange multipliers whose equations of motion enforce the constraints:
\begin{eqnarray}\mbox{3D-Diffeomorphism constraint:}\qquad
\label{mom_ctraint} H_a (x)&=& -2 \,\nabla_b p^b{}_a(x) \approx 0 \qquad \forall x\in M
\\
\label{Ham_ctraint}\mbox{Hamiltonian constraint:}\qquad
 H^\perp(x) &=&  \left( {\sfrac 1 {\sqrt g}}  \left( p^{ab} p_{ab}  - {\sfrac 1 2} (\text{tr} \, p)^2 \right)     - \sqrt g \, R\right)  \approx 0 \qquad \forall x\in M
\end{eqnarray}
The first constraint generates spatial re-labeling of points.  The second constraint is taken to generate local time translations --- refoliations.\footnote{This is a slight exaggeration.  In fact, refoliations have a well defined Hamiltonian representation only for space-times that satisfy Einstein equations \cite{Wald_Lee}.}

\begin{figure}
\begin{center}
 \includegraphics[width=9cm]{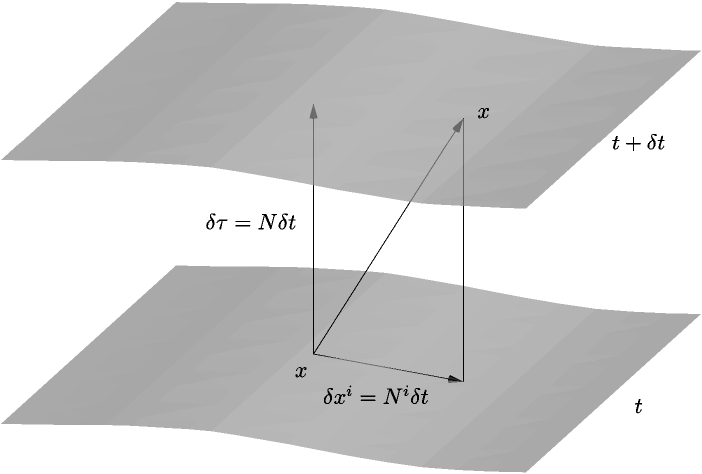}
\caption{A 3+1 decomposition of a globally hyperbolic spacetime. The foliation is built from constant $t$ surfaces. $x$ denotes the spatial coordinates of $M$.}
\end{center}
\end{figure}

  The \lq{}York method\rq{} \cite{York, York2, York3, Niall_York} is the only known mathematical tool for attacking the initial value problem of general relativity \eqref{mom_ctraint},\eqref{Ham_ctraint} generically \cite{3+1_book}. Through this method, the constraints $\mathcal H \approx 0 $ and $\mathcal D_a \approx 0  $ {decouple} and turn into {elliptic equations.} In a closed spatial hypersurface, we start with a reference metric $g_{ab}$, a tensor density $p^{ab}$ and a real number ${\tau}$.  The York method then consists in the following steps:  
  
\begin{flushleft}
1. Given a traceless tensor, one can also find its projection to a transverse component. That is, upon the substitution $p^{ab}\rightarrow p^{ab}-\nabla^{(a} {\chi^{j)}} $ (where round brackets signify index symmetrization, $A_{(ab)}=A_{ab}+A_{ba}$) one can solve the transversality condition (wrt to the vector field ${\chi^a}$):
$$
\nabla_b (\nabla^{(a} {\chi^{b)}} - {\sfrac 2 3} g^{ab} \nabla_c{\chi^c}) = \nabla_b(p^{ab} -{\sfrac 1 3} \text{tr} \, p ~ g^{ab}) ~,
$$

Thereby we obtain the transverse-traceless momentum ${p_{\mbox{\tiny TT}}^{ab}} $:
$$
{p^{ab}_{\mbox{\tiny TT}}} = (p^{ab} -{\sfrac 1 3} g^{ab} \text{tr} \, p) - (\nabla^a {\chi^b} +\nabla^b {\chi^a} - {\sfrac 2 3} g^{ab} \nabla_c {\chi^c})\,,
$$
This means that for any constant $\tau$, the momentum $\pi^{ab}={p^{ab}_{\mbox{\tiny TT}}}+{\sfrac 1 2} {\tau} \, \gamma^{ab} \, \sqrt{\gamma} $ satisfies the constraint \eqref{mom_ctraint}. Importantly, it represents a \emph{CMC slicing} (constant mean curvature), since its trace is necessarily constant:
$$
 g_{ab} \pi^{ab} = {{\sfrac 3 2} \tau} \sqrt{g} = \text{\it const.}~\sqrt{g} ~,
$$

2.  Upon \textit{conformally transforming}  the canonical variables, as $(g_{ab}, {p^{ab}_{\mbox{\tiny TT}}})\rightarrow (\Omega^{-4} g_{ab}, \Omega^4{p^{ab}_{\mbox{\tiny TT}}})$, from the scalar constraint \eqref{Ham_ctraint} one obtains the Lichnerowicz--York equation (LY) wrt the scalar field ${\Omega}$:
$$
 \frac {{\Omega^{-6}}} {\sqrt g}{p^{ab}_{\mbox{\tiny TT}}}^2-{\sfrac 3 8} \sqrt g \, {\Omega^6} {\tau^2} - \sqrt g \left( R \,{\Omega^2} - 8 \, {\Omega} \Delta {\Omega} \right)  = 0~,
$$
This is an elliptic equation which can, under certain additional assumptions,  be solved for $\Omega$, thereby satisfying all of the ADM constraints on a CMC slice $\left( \text{tr} \, {\pi}  = {\sfrac 3 2} \, {\tau}  \,{ \sqrt{{\gamma}}}\right)$.
\end{flushleft}

One important point hidden in the analysis is that  the LY equation and the transversality condition
are conformally invariant:
$$
\left\{\begin{array}{l}
g_{ab} \to {\phi^4} \, g_{ab}\\
p^{ab} \to {\phi^{-4}} \, p^{ab}
\end{array}\right.
 ~~~~ \Rightarrow ~~~~
 \left\{\begin{array}{l}
\gamma_{ab}[{\phi^4} g, {\phi^{-4}} p] = \gamma_{ab}[g, p]\\
\pi^{ab}[{\phi^4} g, {\phi^{-4}} p] = \pi^{ab}[g, p]
\end{array}\right.
$$
This raises an important question: does spatial conformal invariance have any fundamental meaning in general relativity?  In \cite{SD_construction}, we have shown that no other symmetry can be used in this manner to solve the initial value problem of general relativity.  Is this merely a mathematical curiosity or does it carry some insight into gravitational physics?

The natural place to look for the utility and meaning of this symmetry, is to investigate its role in the places where GR fails. By formulating a theory \cite{SD_first, SD_linking} (see also \cite{Flavio_tutorial} for a modern full account) which embodies these symmetries and matches general relativity (wherever the latter admits a global CMC foliation) This has been done to a large extent for black holes \cite{SD_birkhoff, Thin_shell, Flavio_single, Flavio_compact, Flavio_tutorial, Gabe_parity}. Here I want to take a broader look at how one might motivate these symmetries from another place where GR \lq{}fails\rq{}: quantum gravity.

\subsection{Motivation from Quantum Gravity.}
%\paragraph{Renormalization and Lorentz invariance}
\paragraph*{Covariant quantum gravity.} The main technical obstacle to obtaining a theory of quantum gravity, well-documented in the literature, lies in its renormalization properties. Gravity is a non-linear theory, which means that geometrical disturbances around a flat background can act as sources for the geometry itself. Unlike what is the case in other non-linear theories,  GR has the wrong sign of  mass dimension of the gravitational coupling constant, making it naively perturbatively non-renormalizable around Minkowski.\footnote{In fact, the vacuum theory is 1-loop renormalizable, but acquires (cascading) counter-terms at 2-loops --- a fate it encounters at 1-loop by adding sources.} %the `charges'  carried by the non-linear terms in linearized general relativity become too `heavy', generating a cascade of ever increasing types of interactions once one goes to high enough energies. 

But there are other, fundamental problems. For instance, there exists a conflict between our use of quantum mechanics to calculate transition amplitudes from one observable quantity to another,  and the gauge-symmetries of GR. In quantum mechanics, we have time-evolution operators $e^{-i\hat H t}$, taking us from an initial physical state to a final one. In the language of path integrals, it is more convenient to express evolution in terms of a propagator, $K(\phi_1|\phi_2)$, where $\phi$ (e.g. $\phi=(x,t)$) is  deemed gauge-invariant, or in Dirac\rq{}s terminology,  \lq{}observable\rq{}.\footnote{Indeed, the inversion of the quantum mechanical propagator requires gauge-degrees of freedom to have already been gauge-fixed. } 

  In GR, gauge-symmetries are spacetime diffeomorphisms, and the gravitational degrees of freedom that are (non-perturbatively) gauge-invariant thus can correspond to the entire spacetime, e.g. 
 $$\mbox{Diff-invariant quantities:} ~~~\int d^4x \sqrt{{}^\st{(4)}g} {}^\st{(4)}R,~~~  \int d^4x \sqrt{{}^\st{(4)}g}\,~ {}^\st{(4)}R^{\mu\nu}\,\, {}^\st{(4)}R_{\mu\nu}\,\,,~\mbox{etc}$$
Naively combining the two properties would require us to associate a quantum \lq\lq{}time-evolution\rq\rq{} between two different \textit{complete spacetimes}, not between observables within time.\footnote{This argument assumes that spacetime does not possess boundaries, where the gauge-symmetries are fixed. If this is the case, other types of questions arise \cite{Donnelly_2016, Donnelly_local}.} For instance, the fundamental transition amplitude would be of the form $K([{}^\st{(4)}g_1], [{}^\st{(4)}g_2])$, where the square-brackets $[{}^\st{(4)}g]$ signify the quotient of the spacetime metric by all possible gauge-transformations (the spacetime geometry corresponding to the 4-metric ${}^\st{(4)}g$).

 Even if one is willing to accept such fundamental quantum transition amplitudes, there is another problem we must face: there is no known local parametrization of physical (observable) Lorentzian 4-geometries, $[{}^\st{(4)}g]$.\footnote{ Global obstructions --- e.g. the Gribov problem \cite{Singer_Gribov} --- also exist, but are less concerning.}  To understand what this means, we need to introduce the notion of a \lq{}slice\rq{}. A \lq{}slice\rq{} is a split between the physically equivalent (or gauge-equivalent) field configurations and the physically distinct ones. A local slice is one that performs this split only locally in field space, and it is equivalent to a local gauge-fixing if gauge-transformations form well-defined gauge-orbits (see figure 2).  Finding a slice theorem is relatively straightforward in the case of metrics of Euclidean signature (for any dimension). This is shown in appendix \ref{app:slice}.
 \begin{figure}\begin{center}
\includegraphics[width=.4\textwidth]{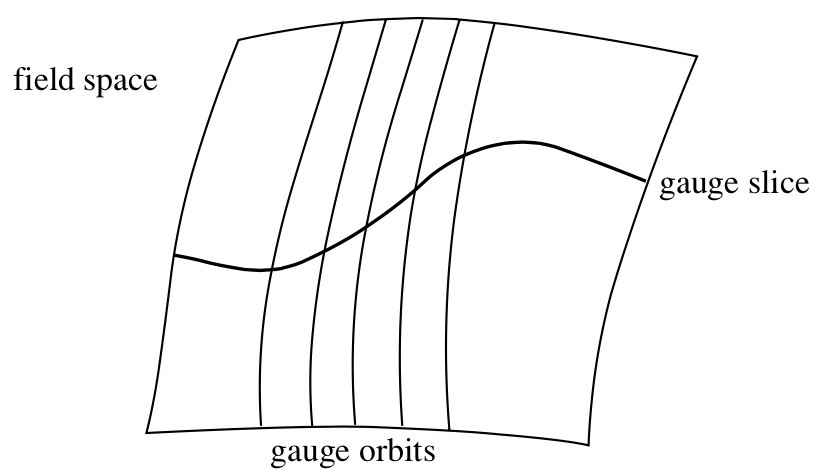}
\end{center}
\caption{A \lq{}gauge-slice\rq{} in the field-space of a field theory with gauge-symmetry.}
\end{figure}

In sum, what goes wrong in the Lorentzian case involves the lack of invertibility of certain second order differential operators (given in equation \eqref{elliptic} in appendix \ref{app:slice}). To slightly expand, in the Euclidean case,  defining a supermetric on field space, one can find orthogonal directions $T_g^\perp\mathcal{O}_g$ to the orbits $\mathcal{O}_g$ of the metric $g$ under the action of the diffeomorphisms of $M$, Diff$(M)$. Such transversal directions to the gauge orbits involve finding the adjoint space of the Killing form, $\mathcal{L}_\xi g_{\mu\nu}$, which requires the orthogonality conditions to be described by elliptic operators. In the case of Lorentzian metrics, the analogous operators are hyperbolic, invalidating the construction of the slice. It is the sign of Time that gets in the way.

  A local parametrization, or slice, has only been constructed  (by Isenberg and Marsden) for Lorentzian metrics which i) satisfy the Einstein equations and ii) which furthermore admit a particular kind of time-foliation \cite{IsMa1982}.\footnote{ Since we do not know how to use the standard techniques in the case of Lorentzian metrics, the strategy is to reduce the symmetries to those acting on a spatial slice.  To have a correspondence between initial data and a full spacetime geometry, one needs  uniqueness of  maximal extensions of  spacetime developments from initial data \cite{Hawking_Ellis}. Thus  the restriction to the Einstein equations, whereby we can restrict attention to $(\gamma, \pi)\in \mathfrak{C}(M)$ where $\mathfrak{C}(M)$ is the subspace of phase space satisfying the Einstein constraint equations on the manifold $M$. Although $\mathfrak{C}(M)$ is not a manifold (essentially because of linearization instabilities), the intersection of $\mathfrak{C}(\Sigma)$ with the subspace tr$\pi=$const., is a regular manifold \cite{Fischer_Marsden}. One can then use uniqueness of CMC foliations to fix the time-like component of space-time diffeomorphisms. Identifying initial data with CMC data, one now uses a slice theorem of the spatial (or foliation-preserving) diffeomorphisms acting on $\mathfrak{C}(M)\cap\, \{$  tr$\pi=$const$\}$. }    This choice ---  Constant Mean Curvature (CMC) --- corresponds to synchronizing clocks  so that they measure the same expansion rate of space everywhere (i.e. same local Hubble parameter).   They also correspond to the foliation required by the York method, as seen above.

There are thus two problems touched on above: 
i) There is no known construction of a slice for the space of Lorentzian metrics. It is thus not clear if one can identify gauge degrees of freedom and get to a parametrization of observables. For on-shell metrics (which moreover admit a CMC foliation), one can indeed find a slice; but not allowing off-shell metrics to be represented makes it ill-suited for quantum gravity. ii) Even ignoring this issue (by for example considering Euclidean quantum gravity), it is not obvious how we should interpret a transition amplitude $K([{}^\st{(4)}g_1], [{}^\st{(4)}g_2])$ as an evolution in the standard, predictive scientific framework, which associates a transition probability between two \lq\lq{}instantaneous\rq\rq{} state of affairs. Indeed, even in  Yang-Mills theory, non-local spatial properties (arising from gauge-invariance arguments) are treated differently than non-local temporal ones \cite{Lavelle_review}.

\paragraph*{Canonical quantum gravity and the Problem of Time}

One way to overcome this issue is to a priori base the theory not on an underlying covariant spacetime, but on a more dynamical account. It turns out that indeed, such breaking of covariance can also help with renormalization properties, as is the case with Horava-Lifshitz \cite{Horava, Barvinsky_hor}. Of course, these theories have other problems;  they often carry scalar gravitational degrees of freedom due to their reduced symmetry content.

{However,  for the dynamical account of GR --- or canonical, or Hamiltonian, framework --- the required slicing of spacetime into equal-time surfaces  is merely an auxiliary structure, and  physics must be independent of such choices.
 Indeed, the canonical system comes with a conserved quantity ---the Hamiltonian constraint seen in equation \eqref{Ham_ctraint}--- which has an associated symmetry transformation, and that gives us back the freedom of choosing such artificial slicings.} %Let me explain: analogously to Noether\rq{}s theorem,\footnote{Basically, Noether\rq{}s theorem states that associated to each conserved quantity is a symmetry.  I only deal with the version of the theorem concerned with local symmetries, i.e. not associated to physical charges. } this constraint --- essentially a conserved quantity --- generates a freedom to  \lq{}refoliate\rq{}  space-time, re-defining the surfaces of constant time.
   In this way, local time translations become part of the symmetries of the theory.  Since  physical observables should not vary under the action of a symmetry transformation, they are required to be non-local in time.%\footnote{ } This is the source of the Problem of Time (see e.g. \cite{Kuchar_time, Isham_POT}).
   
   The hope is then that one can single out the physical instantaneous {degrees-of-freedom} which one would like to quantize, and then evolve in a more standard quantum mechanical framework. 
 It would be convenient if quantum evolution could correspond to a transition between instantaneous configuration {degrees-of-freedom} also for gravity.   The problem here is that  the ADM configuration space  --- the space of Riemannian 3-metrics --- does not itself carry an intrinsic representation of the local space-time symmetries.   % because the action of the constraints on the gravitational phase space ($T^*$Riem) does not project down to an intrinsic action on gravitational configuration space (Riem). %This should come as no surprise. The instantaneous physical observables need to correspond to the entire space-time history, and so  require information about the initial rate of change of the spatial metric, i.e. information about the gravitational momenta (which live on $T^*$Riem). 
In other words, as we show in the next section, refoliations don\rq{}t form gauge-orbits in the space of Riemannian 3-metrics, and thus no \lq{}slice\rq{} can be found for them.

%there is no well-defined \lq{}reduced configuration space\rq{} corresponding to the canonical symmetries of GR.\footnote{ For canonical GR, in very special cases with high degree of symmetry --- e.g., the polarized Gowdy spacetimes --- one can isolate the {degrees-of-freedom}. Similarly, if one restricts oneself to globally hyperbolic spacetimes with CMC foliations by 3-dimensional manifolds which only admit Yamabe negative metrics, there is a very formal reduction that can be carried out \cite{Fischer_Marsden}. However, in this later case  the reduction is also  not explicit, and both cases are special. See also footnote \ref{footnote:phys_dofs} for more on this point. } Thus there is no access to the fundamental instantaneous configuration {degrees-of-freedom}, the ones we would like to apply standard quantization to. In the words of the renowned mathematical relativist Jim Isenberg:\begin{quote} \lq\lq{}In general, as far as obtaining an explicit reduced phase space corresponding to the {degrees-of-freedom} of the gravitational field of Einstein\rq{}s theory --- the thing you might want to use as the starting point for a quantization program --- it is fair to say that no one knows how to do this. As well, I\rq{}m guessing that no one will figure out a way to do this explicitly.  \cite{Isenberg_priv}\lq\lq{}\end{quote}

\paragraph*{Maximal symmetries of Riem.}

But what if we go the opposite way: instead of assuming the canonical symmetries associated to general relativity --- spatial diffeomorphisms and local time reparametrizations ---  we investigate what kind of local symmetries still allow a description of the \textit{invariant} --- i.e. physical --- and  \textit{instantaneous} configuration  {degrees-of-freedom}? Such symmetries would be compatible with standard quantum evolution between physically different instantaneous configurations.  

This is how I will start the technical developments of the paper --- by looking at the possible symmetry content such a theory can carry.  %Here I will start from this assumption -- that the theory is based on smooth spatial (i.e. not causally related) fields. I will then show what types of theories are compatible with such an assumption, and some of their surprising characteristics.
In the case of gravity, under some assumptions, it can be shown that the non-trivial allowed symmetries that do act as a group in the metric configuration space  are precisely spatial diffeomorphisms and scale transformations. The remaining physical independent {degrees-of-freedom} --- the quotient of the space of spatial metrics by this full group of local symmetries --- belong to \lq{}conformal superspace\rq{}, the space of conformal spatial geometries. Unlike what is the case for the canonical configuration {degrees-of-freedom} of GR,  this invariant quotient is well-defined and has a simple geometric interpretation: it describes  spatial angles, with two {degrees-of-freedom} per point. %A conformal geometry is the level of structure in-between the merely topological and the fully geometrical. That quantum mechanical evolution should refer to transitions between such simple geometrical {degrees-of-freedom} would elegantly reconcile our experience of Time with our experience of spacetime. 

\begin{figure}[h!]\label{}
\center
\includegraphics[width=0.6\textwidth]{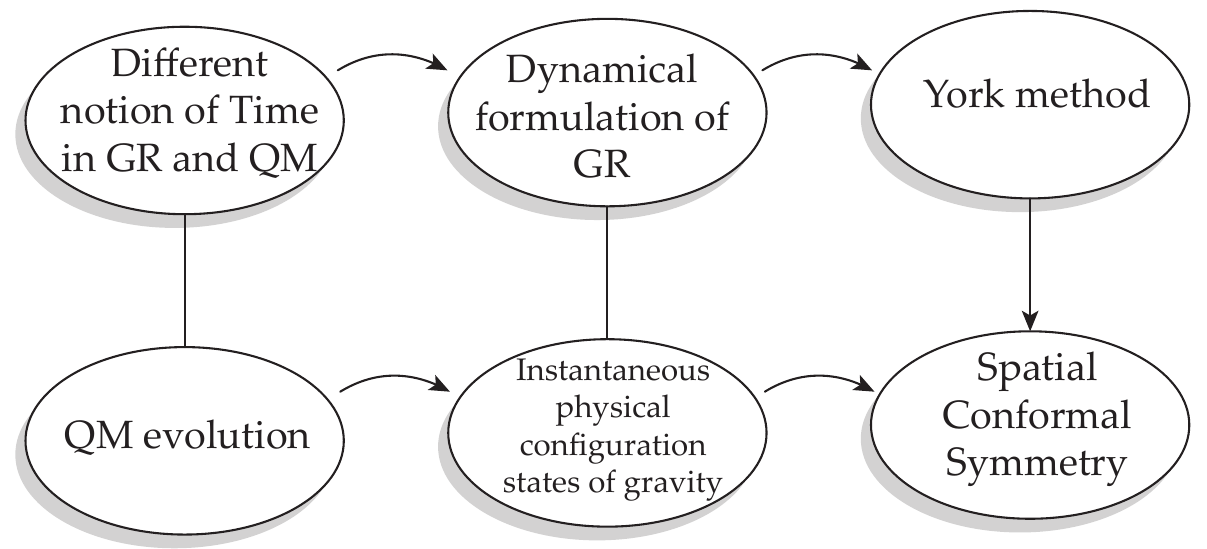}
\caption{The convergent arguments for fundamental spatial conformal symmetry. The top sequence can be seen to describe the need for the York method. In the first part of the paper, I will argue for the bottom sequence of arguments, and explore a few consequences.}
\end{figure}
%$$\begin{array}{rlc} \neq\mbox{time in GR and QM} \Rightarrow  \mbox{Dynamical formulation of GR} & \Rightarrow& \mbox{York method}\\
%~~~~~~~~~~~~~~~~~~~~~~~~ & ~ &~\Downarrow~\\
 %\mbox{QM evolution} \Rightarrow  \mbox{Instantan. physical states} &\Rightarrow&  \mbox{Conformal sym.}
%\end{array}$$
There are thus two lines of converging arguments (see figure 3).    Both the York method and the slice theorem require that spacetime be CMC-foliable. The York method is a tool used to obtain a dynamical description of the metric, and Isenberg and Marsden\rq{}s slice theorem uses CMC slicing to avoid complications due to the negative signs of the Lorentzian spacetime metric.  But CMC foliations have the curious property that, in them,  the gravitational dynamics of the pure spatial scale of the metric decouple from the scale-free, or \lq\lq{}shape\rq\rq{}, degrees of freedom.   I.e. it allows one to describe gravitational evolution in this auxiliary time in terms of spatially \lq{}conformally invariant\rq{} geometries \cite{York, SD_first, Flavio_tutorial}.  
 % Indeed,  most of the formal proofs for existence and uniqueness of solutions of the general relativistic {dynamics} requires the use of this particular choice of time-labels (CMC). 
That is, within the dynamical framework, the phase space variables can be constrained so that they describe only CMC foliations. The constraints they are obligated to satisfy in order to be CMC, tr$(\pi)=$const., as per Noether\rq{}s theorem, generate canonical transformations.  The rightmost downward arrow in figure 3 represents the fact that such transformations are exactly changes of spatial scale, i.e.  the local conformal transformations which are the subject of this paper. %\footnote{In fact, the momenta are set to be transverse and constant trace, not trace-free; which would be the true generator of conformal transformations.  However, one can interpret the constant trace part as defining an auxiliary quantity, the York time. The transverse traceless choice sets the gravitational momenta to be purely tensorial (spin-2), with no scalar (spin-0) or vector (spin-1) components.} 
 In sum, surprisingly,  both the York method \cite{York} and the slice theorem \cite{IsMa1982} show that although GR is not fundamentally concerned with spatial conformal  geometries, it is deeply related to them, furthering the need for the present investigations. 

Indeed, shape dynamics \cite{SD_first, Flavio_tutorial} amounts to taking these hints --- coming from the initial value formulation of general relativity and from the consideration of symmetries compatible with a quantum mechanical evolution between instantaneous physical configuration states --- seriously. It does this by considering dynamical systems whose natural habitat is conformal superspace. However, in the way it was constructed to obtain a very explicit correspondence to GR dynamics, shape dynamics required a non-local Hamiltonian.\footnote{However, see \cite{SD_construction} for a first-principles construction of shape dynamics with no reliance on GR.}  It is an interesting question then, to ask what kind of theory one can obtain merely from demanding  locality and  compatibility with the symmetry content. The existence and properties of such theories is a considerably important point to settle in the entire spatially relational approach to which shape dynamics belongs, and is moreover motivated from the above convergent arguments leading us to consider spatial conformal symmetry in its own right;  even if merely to phenomenologically rule out the other models.

\paragraph*{Roadmap}
%In the remainder of this introduction, I will expand on two obstacles that stand in the way of reconciling, on one hand, the correct number and type of gravitational degrees of  freedom with, on the other,  a quantum transition between instantaneous observables. Two obstacles lie in the way of accomplishing this: i) even abstractly, there is no known parametrization of the  distinct Lorentzian geometries (for closed spatial manifolds), ii) the space of instantaneous configurations of GR ---the space of Riemannian 3-metrics, Riem--- doesn\rq{}t admit a local product structure with respect to the action of spacetime diffeomorphisms. 

In section \ref{sec:symmetries} I will show that conformal diffeomorphisms are the maximal group of symmetries  compatible with the notion of instantaneous evolutions of spatial metrics. In section \ref{sec:past}, I will show that the physical (or reduced) configuration space formed by the quotient wrt these symmetries has an interesting topological structure, which in some cases singles out a unique least dimensional element. In section \ref{sec:form_Lagrangian} I will display the type of local square-root Lagrangian densities compatible with this framework, and include matter couplings in \ref{sec:matter}. Through the use of matter couplings, we can recover an ``experienced duration'' and reconstruct a spacetime from field-histories. We find that static Bainchi IX and (approximate) Schwarzschild are recoverable in this fashion through simple couplings of gravity to its sources. In section \ref{app:conformal_action}, I start the study of quantization of the theory using the unique preferred least-dimensional point in reduced configuration space as an ``anchor'' for constructing the wave-function of the Universe through a path-integral transition amplitude. I show that such a theory has simple BRST symmetries and the wave-function conservation laws. I use a given background to show that the propagator will include higher order contributions, likening properties of the theory with those of Horava-Lifschitz. I then end with the conclusions. 

% The main idea behind a dynamical point of view is to set up initial conditions and construct the spacetime geometry by deterministically evolving in a given auxiliary definition of time. Having a notion of evolution already makes a dynamical setting for gravity more compatible with quantum mechanics and its physical transformations from one physical state to another.  
%I will show that the simplest of these theories, when coupled to other sources in the most naive way,  can give standard notions of space-time. I will exemplify this mechanism with approximate Bianchi IX and Schwarzschild types of solutions.  I will then investigate some of the properties of the quantum theory, such as its BRST structure and its propagator around a conformally flat solution, showing that it has some of the desirable properties of Horava-Lifschitz in this respect.
 
\section{The Classical Theory}\label{sec:observers}
Here I will show how to construct the most general action compatible with local symmetries in configuration space. 
\paragraph*{Basic structure of Riem.}
In the present context, I will let $M$ denote a spatial, closed (i.e. compact without boundary)  $3$-dimensional manifold. 
For the non-relativistic gravitational systems in consideration here, I will take configuration space to be the space Riem$(M)=:\mathcal{Q}$, of positive-definite sections of the symmetric covariant tensor bundle $C_+^\infty(T^*M\otimes_S T^*M)$ over $M$, which forms a subspace (a cone) of the Banach vector space $\mathbb{B}:=C^\infty(T^*M\otimes_S T^*M)$.\footnote{Here I have chosen the degree of differentiability to be infinity, i.e. smooth functions. I will need this assumption in order to prove the uniqueness of conformal symmetry; I need to consider functionals with an arbitrary number of derivatives in the metric. However, it is well-known that in this case the space of sections above will form instead an inverse limit Banach manifold (Hilbert, once we have introduced a complete metric) by the Sobolev construction. I will ignore these more technical obstructions here, assuming they do not interfere with what I would like to achieve. For more on these matters, see \cite{Gauge_riem, Ebin, FiMa77}.} This subspace has a one-parameter family of natural Riemannian structures, induced pointwise by the metric $g_{ab}$:
\be\label{equ:supermetric} (v, w)_g:=\int d^3 x\sqrt{g}\, G_\lambda^{abcd}v_{ab}w_{cd}
\ee
where $G_\lambda^{abcd}:=g^{ac}g^{bd}-\lambda g^{ab}g^{cd}$ (when acting on symmetric tensor fields, $v_{ab}=v_{(ab)}$ and so on), and $0\leq\lambda< 1/3$. 
These are called the DeWitt supermetrics (with DeWitt value $\lambda$), and they are often useful for explicit calculation, such as in the proof of  slice theorems \cite{Ebin, FiMa77}.

\subsection{Finding the most general symmetries compatible with our framework.}\label{sec:symmetries}
If one takes configuration space to be fundamental, one must consider symmetries that act intrinsically on it. 
Here, I will argue that the most general such symmetries are conformal diffeomorphism transformations. 

To find them, we first look at the Hamiltonian vector field associated to a  smeared functional $F[g,\pi,\lambda]$, polynomial in its variables. These functions generate transformations in canonical variables through the action of their Poisson bracket, i.e. they have vector field $\chi_F$ generating flow along phase space as in $\imath_{\chi_F}\Omega=d F$, where $\Omega$ is the canonical symplectic structure, and $\imath$ is the inner derivative. See figure 4. 
\begin{figure}[h!]\label{}
\center
\includegraphics[width=0.5\textwidth]{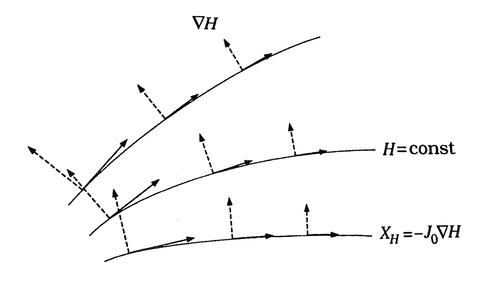}
\caption{Relation between a scalar function and its associated symplectic vector field. Here $J_0$ is the symplectic inversion map, and $\nabla H$ is the gradient of the scalar function $H$, whose regular values define surfaces in phase space. }
\end{figure}
For example, in general relativity, the  scalar (super-Hamiltonian) constraints are given by \eqref{Ham_ctraint}, which we reproduce here: 
\be\label{Ham_ADM}H^\perp(x):=\left(\frac{\pi^{ab}\pi_{ab}-\frac{1}{2}\pi^2}{\sqrt{g}}-R\sqrt{g}\right)(x)=0\ee
and, given a particular linear combination of these constraints, defined by a smearing $\lambda^\perp$, it generates the following transformation:
\be\label{refol_sym}\delta_{\lambda^\perp}g_{ab}(x)=\frac{2\lambda^\perp (\pi_{ab}-\frac12 \pi g_{ab} )  }{\sqrt{g}}(x)\ee depends not only on the metric, but also on the momenta (see figure 5). 
 \begin{figure}[h!]\label{}
\center
\includegraphics[width=0.7\textwidth]{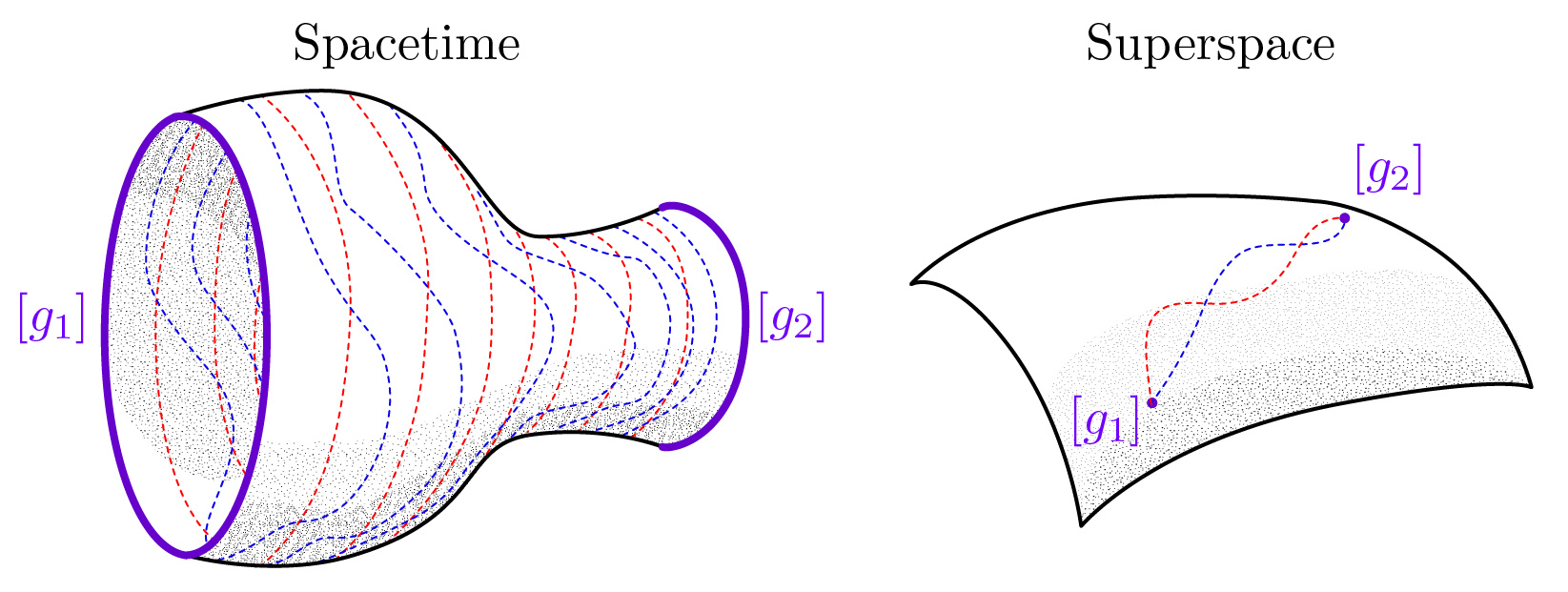}
\caption{Many-fingered time issue: a single solution to Einstein's equations
(the slab of spacetime on the left) between two Cauchy hypersurfaces (in purple)
where two initial and final 3-geometries $[g_1], [g_2]$ (the [ ] brackets stand for
\lq{}equivalence class under diffeomorphisms') are specified, does not correspond to
a single curve in superspace (on the right). Instead, for each choice of foliation
of the spacetime on the left that is compatible with the boundary conditions
$[g_1], [g_2]$ there is a different curve in superspace. Here I have represented two
different choices of foliation in red and  blue. The point being that one cannot find the \lq\lq{}superspace\rq\rq{} equivalent to the full set of local symmetries of GR. This section shows that the maximal local symmetry acting on the space of 3-metrics for which this is possible is conformal diffeomorphisms, giving conformal superspace. Figure taken from arxiv \cite{Flavio_tutorial} with permission.}
\end{figure}
As a counter-example, the spatial diffeomorpshisms \eqref{mom_ctraint}, generated by $H^a=-\nabla_b \pi^{ab}=0$,  act with the smearing $\lambda^a$ as:
$$\delta_{\vec{\lambda}}g_{ab}(x)=\mathcal{L}_{\vec{\lambda}}g_{ab}(x)$$
which has pointwise  dependence  in $\mathcal{Q}$ (and $M$).

  Indeed, for  the associated symmetry to have an action on configuration space that is independent of the momenta, $F[g,\pi,\lambda]$ must be linear in the momenta. This already severely restricts the forms of the functional to 
\be\label{equ:total_symmetry}F[g,\pi,\lambda]=\int \tilde F(g,\lambda)_{ab}(x)\pi^{ab}(x)\ee
so that the infinitesimal gauge transformation for the gauge-parameter $\lambda$ gives:
$$\delta_\lambda g_{ab}(x)=\tilde{F}(g,\lambda)_{ab}(x)$$
A Poisson bracket here results in
\be\label{equ:Poisson_bracket}
\{F[g,\pi,\lambda_1],F[g,\pi,\lambda_2]\}=\int d^ 3x  \left(\frac{\delta \tilde{F}(g,\lambda_1)_{ab}}{\delta g_{cd}}\pi^{ab}\tilde{F}(g,\lambda_2)_{cd}-\frac{\delta \tilde{F}(g,\lambda_2)_{ab}}{\delta g_{cd}}\pi^{ab}\tilde{F}(g,\lambda_1)_{cd}\right)
\ee
and this must close in order that it has any chances of being a symmetry generator (i.e. it must be a first class constraint). 

Let us for now assume that $\tilde{F}(g,\lambda)_{ab}$ is covariant tensor of rank two. This possibly requires integrating away covariant derivatives from $\pi^{ab}$.  
If $\tilde{F}$ has no derivatives of the metric, $F[g,\pi,\lambda_1],F[g,\pi,\lambda_2]$  will straightforwardly commute. But with no derivatives the only objects we can form are: 
$$ \tilde{F}(g,\lambda)_{ab}=\lambda g_{ab}\, ~,~\mbox{and} \qquad \tilde{F}(g,\lambda)_{ab}=\lambda_ {ab}
$$
In the first case, $F$ generates spatial conformal transformations, since:
$$\{F[g,\pi,\lambda], g_{ab}(x)\}=\lambda(x)g_{ab}(x)~~,\qquad \{F[g,\pi,\lambda], \pi^{ab}(x)\}=-\lambda(x)\pi^{ab}(x)$$
  In the second case, they would imply that $\pi^{ab}=0$, a constraint killing any possibility of dynamics.

 Suppose  $\tilde{F}$ is a tensor with two or more derivatives of the metric. For example,  
$$ \tilde{F}(g,\lambda)_{ab}= \lambda(\alpha R_{ab}+\beta R g_{ab})
$$
A substantial amount of computations show that the algebra  of equation \eqref{equ:Poisson_bracket} does not close for any values of $\alpha$ and $\beta$. To see this, one needs to first calculate the general bracket, integrate away derivatives from one of the smearings, $\lambda_1$ which is then set to a Dirac-delta function, $\lambda_1(x)=\delta(x,y)$. This gets rid of the integral signs and gives us a local density which must vanish, for all values of the other smearing function, $\lambda_2(x)$. To finish the proof, one shows that there are terms that cannot be proportional to $F[g,\pi, \lambda]$.

As a specific example, the Ricci flow, or the conformal Ricci flow do not have a local group action in Riem, in the sense that commuting the flow with respect to two different smearings (usually the Ricci-flow is taken with respect to a unit smearing), will not give another Ricci-flow. That is, the local generator of Ricci-flow on phase space is $\tilde{F}(g,\lambda)_{ab}(x)=\lambda R_{ab}(x)$, since this is what will generate $\delta_\lambda g_{ab}(x)=\lambda R_{ab}(x)$. Performing the algorithm described above (see auxiliary Mathematica file), one obtains, integrating away all derivatives away from $\lambda_2$ and setting it as a Dirac delta, a single term containing derivatives of the momenta: 
$$\nabla^a\lambda_1 (2  R^{bc} (\nabla_a\pi_{bc} -\nabla_c\pi_{ab}))$$ 
which therefore cannot vanish by combination with the other terms.\footnote{Unless we further stipulate symmetries, which in this case would be overly restrictive in any case.}

    If one instead chose a term of the form $\beta R\lambda_{ab}$ one can show that the rank of this constraint is not constant along phase space, and moreover it would imply that $\pi^{ab}=0$ almost everywhere, as before.  
Taking \eqref{equ:total_symmetry} to contain covariant derivatives of $\pi_{ab}$, with terms  of the form e.g.: $\lambda\nabla_cR\nabla_c\pi^{ab}$, is tantamount to taking a generator of the form $\lambda \nabla^2 R +\partial_c\lambda \nabla^c R$. It can be shown that  only the momentum constraint, generator of diffeomorphisms,  $F(\lambda, g,\pi)=\int d^3 x\, \lambda_a\nabla_b\pi^{ab}$ survives as the generators of a closed algebra \cite{SD_construction}.\footnote{In fact,  the constraint $(\nabla^2-\frac{1}{8}R)\pi=0$ can also survive. This is the conformal Laplacian in 3-dimensions acting on the trace of the momentum (which is a conformally invariant scalar density). By exploring conformal transformations one can show that this reduces to the conformal constraint $\pi=0$ almost everywhere on phase space as well \cite{SD_construction}. }

These conclusions hold order by order in number of derivatives of the metric (see \cite{Vasu_rigidity} for a method of proof that would need to be pursued here, and  accompanying auxiliary Mathematica file for the calculations).   The outcome  is that the only relevant geometrical structure between the differentiable one and the metric one is conformal geometry. 

Based on these considerations,  I will take the gauge symmetries in Riem$(M)=\mathcal{Q}$ compatible with the foundation of the theory to be at most conformal diffeomorphism transformations.

\subsection{The structure of conformal superspace.}\label{sec:past}

 Theories with gauge symmetry are defined by possessing redundant descriptions. In the present context, this refers to 3-metrics, which are related by a symmetry transformation. Usually, a symmetry is only defined once we have an action functional, by those transformations which leave such a functional invariant. Here I am pursuing a different strategy: defining the local symmetries by demanding that configuration space have a well-defined quotient. The appropriate mathematical tool for defining and exploring such a quotient is called \lq\lq{}a slice\rq\rq{}.  Its use shows us that the reduced configuration space has a rich geometrical structure \cite{Fischer, FiMa77}; a structure we can co-opt in order to build a wave-function of the Universe, as e.g. in the no-boundary proposal \cite{Hartle_Hawking}.  The difference is that unlike in Hartle-Hawking, due to the physical nature of reduced configuration space, we can obtain boundary conditions for the wave-function directly from gauge-invariant variational principles, i.e. they are less arbitrary.

\paragraph*{Basic structure}
  Conformal superspace is the name given to the quotient of configuration space with respect to the group of conformal transformations. First, let\rq{}s explain a bit further what this group is. I will here mostly follow the nomenclature of \cite{FiMa77}.

Spatial diffeomorphisms, $f\in{\mbox{Diff}}(M)$,  act on the metric through pull-back, ${\mbox{Diff}}\times\mathcal{Q}\ni (f, g)\mapsto f^*g\in \mathcal{Q}$. The infinitesimal action is given by the Lie derivative, i.e. for $\xi^a$ the vector field flow of $f$,  $\frac{d}{dt}(f_t^*g)=\mathcal{L}_\xi g$, where $f_0=$Id. The orbits of the spatial diffeomorphisms in $\mathcal{Q}$ are along Killing directions wrt to the metric \eqref{equ:supermetric} (see \cite{DeWitt_QG1}).
Conformal transformations are designated by $\mathcal{P}$, the group of positive scalar functions on $M$, acting pointwise through multiplication, $\mathcal{P}\times\mathcal{Q}\ni (\rho, g)\mapsto \rho g\in \mathcal{Q}$.

With these two groups, we can form the semi-direct product $\mathcal{C}:={\mbox{Diff}}\ltimes \mathcal{P}$, acting on $\mathcal{Q}$. As ${\mbox{Diff}}$, the set of conformal diffeomorphisms $\mathcal{C}$ forms an infinite-dimensional Lie group, with Lie algebra given by the semi-direct sum of smooth vector fields $C^\infty(TM)$ and scalar functions $C^\infty(M)$. That is, let, $f_1,f_2$ be diffeomorphisms, and $\rho_1, \rho_2$ two scalar transformations, conformal diffemorphisms have
group structure $(f_1,\rho_1)\cdot(f_2,\rho_2)=(f_1\circ f_2,\rho_2(\rho_1(f_2)))$ where $\rho_2(\rho_1(f_2))$ just means
scalar multiplication at each $x\in M$ as $\rho_2(x)(\rho_1(f_2(x)))$. As with ${\mbox{Diff}}$, thus $\mathcal{C}$ is an
infinite-dimensional regular Lie group and it acts on $\mathcal{Q}$
 on the right as a group of transformations by:
\begin{eqnarray*}
 (Diff\ltimes \mathcal{P})\times \mathcal{Q} &\rightarrow& \mathcal{Q} \\
~((f,\rho),g)&\mapsto & \rho (f^*g)
\end{eqnarray*}

\paragraph*{The slice theorem and stratified manifolds.}
Roughly speaking, a slice for the action of a group $G$ on a manifold $\mathcal{X}$ at a point $x\in \mathcal{X}$ is a manifold $S_x$, transversal to the orbit of $x$,  $\mathcal{O}_x$ (see appendix \ref{app:slice} for a rough and brief sketch of definitions, corollaries and how proofs of existence work).  If the isotropy group $G_x$ of $x$ is trivial, then $S_x$ gives a local chart for the space $\mathcal{X}/G$ near $x$. I.e. it gives $\mathcal{X}$ a local product structure, with one of the factors being isomorphic to the group, $\mathcal{U}\simeq S_x\times G$, with $\mathcal{U}$ being a proper subset of $\mathcal{X}$ containing $x$ (see corollary \ref{cor} in appendix \ref{app:slice}). In this case, one can use the slice to parametrize the physically distinct configurations. But complications arise when the isotropy groups of $x\in \mathcal{X}$ become non-trivial, because the symmetry group in question may act qualitatively differently on different orbits.

 In that case,  let $\mathcal{N}_{x_o}=\{x\in \mathcal{X}~|~I_x ~~ \mbox{is conjugate to}~~ I_{x_o}\}$. Then according to corollary \ref{cor}, a slice theorem  shows that $\mathcal{N}_{x}/G$ is a manifold (since $I_x$ doesn\rq{}t change dimension). Each such manifold defines a  \lq{}stratum\rq{},  containing  the orbit $\mathcal{O}_x$. Then (see \cite{IsMa1982} for a review):
 \begin{cor}[Stratification]
There exists an isomorphism $\mathcal{X}/G=\mathcal{S}_1\cup \mathcal{S}_2\cdots \mathcal{S}_n$ 
with $\mathcal{S}_i$ a stratum as above,  where $i$ characterizes the dimensionality of an isotropy group, with 
$\mathcal{S}_1=\hat{\mathcal{X}}/G$  
(for $\hat{\mathcal{X}}=\{x\in \mathcal{X}~|~I_x=Id\}$), and $\mathcal{S}_{i+1}\subset \partial \mathcal{S}_i$. I.e. the strata are ordered by increasing symmetry and decreasing dimensionality. 
 \end{cor}

 Let us take the diffeomorphism group ${\mbox{Diff}}=$Diff$(M)$, acting on the metrics through pull-back $f^*g_{ab}$, $f\in{\mbox{Diff}}$. For metrics that have non-trivial isometry groups, $I_g\in{\mbox{Diff}}$, we have a degenerate action of the diffeomorphism group. For this reason,  the quotient wrt the diffeomorphisms of the space of metrics  $\mathcal{Q}$ over $M$,  has  a structure with different strata. 

  Stratified manifolds have nested \lq\lq{}corners\rq\rq{} -- each stratum corresponds to a dimension of the stabilizer group, and  has as boundaries a lesser dimensional stratum.  The larger the stabilizer group, the lower the strata. Let $\mathcal{Q}_o$ be the set of metrics without isometries. This is a dense and open subset of $\mathcal{Q}$, the space of smooth metrics over $M$. Let  $I_{n}$ be the isometry group of the metrics $g_n$, such that the dimension of $I_n$ is $d_n$. Then the quotient space of metrics with isometry group $I_n$ forms a manifold with boundaries, $\mathcal{Q}_n/\mbox{Diff}(M)=\mathcal{S}_n$. The boundary of $\mathcal{S}_n$ decomposes into the union of $\mathcal{S}_{n\rq{}}$ for $n\rq{}>n$ (see \cite{Fischer}). 
  
  A useful picture to have in mind for this structure is  a \lq\lq{}bottomless\rq\rq{} tetrahedron (seen as a manifold with boundaries). The interior of the tetrahedron has boundaries which decomposes into faces, whose boundaries decompose into lines, whose boundaries are the single vertex at the top. The single vertex at the top of the tetrahedron is geometrically singled out, and we will use it to construct our path integral.

  \subsection{The form of the Lagrangian}\label{sec:form_Lagrangian}
  
  Now that we have explored the gauge structure of the symmetries which have the appropriate action on metric configuration space, $\mathcal{Q}$, we move on, to calculate the possible form of local square-root Lagrangians which possess such symmetries. 
 
  To start the calculation, we note that if the 3-metric $g_{ab}$ has conformal weight $4$, i.e.$\delta_\epsilon g_{ab}= {4\epsilon} g_{ab}$,  the symmetric 2-tensor $\dot{g}_{ab}$, also  has conformal weight $4$, and the undensitized totally anti-symmetric 3-tensor $\epsilon^{abc}$ has conformal weight $-6$.  

Thus, as a necessary condition  to have a conformal diffeomorphism invariant action, we must match the following tensor  indices and conformal weights:
\begin{eqnarray}
\mbox{Tensor:}&\qquad -3N_\epsilon+ 2N_{\dot g}+2N_g-2N_{g^{-1}}+N_{\nabla}&=0\label{equ:counting_tensor}\\
\mbox{Conformal:}&\qquad -3N_\epsilon +2N_{\dot g}+2N_g-2N_{g^{-1}}&=-3\label{equ:counting_conf}\\
\nonumber\Rightarrow& N_\nabla= -3 &~ 
\end{eqnarray}
It turns out that the only polynomial invariant we can form with three derivatives of the metric is the Chern-Simons functional \cite{Bertlmann},
\be \label{equ:CS}CS[g]= \int d^ 3 x (d\Gamma\wedge \Gamma+\frac{2}{3}\Gamma^ 3)
\ee 
where $\Gamma(g_{ab})$ is the  Levi-Civita  connection one form associated to $g_{ab}$. 

Since $CS[g]$ is a conformal diffeomorphism invariant, From functionally differentiating it, we obtain the symmetric tensor: 
$$\frac{\delta{CS[g]}}{\delta g_{ab}}=\sqrt{g}C^ {ab}$$
where the (undensitized) Cotton tensor is defined as:
\be\label{equ:cotton_tensor} {C}^{ab}:={\epsilon}^{acd}\nabla_c\left({R^b}_d-\frac{1}{4}\delta^b_d R\right)
\ee
where here we are using the undensitized totally anti-symmetric pseudo-tensor $\epsilon^{abc}$. 

Since it is a functional derivative of a conformal diffeomorhism invariant functional, under an infinitesimal diffeomorhism $\delta_\xi g_{ab}=\mathcal{L}_\xi g_{ab}$,  or infinitesimal conformal transformation, $\delta_\rho g_{ab}=\rho g_{ab}$, it must remain invariant. Thus besides being symmetric, we obtain equations telling us that the Cotton tensor is also transverse and traceless:
\be\label{equ:cotton_tensor_props} C^{ab}=C^{(ab)}~~,~~ {C^{ab}}_{;b}=0~~,~~{C^a}_a=0
\ee

Thus, under a conformal transformation  $g_{ab}\rightarrow e^{4\rho}g_{ab}$
$$\sqrt{g}e^ {6\rho}C^ {ab}[e^ {4\rho}g]= \frac{\delta{CS[e^ {4\rho} g]}}{\delta (e^{4\rho}g_{ab})}=e^{-4\rho} \frac{\delta{CS[g]}}{\delta g_{ab}}=e^{-4\rho}\sqrt{g}C^ {ab}[g]$$
 we get $C^{ab}\rightarrow e^{-10\rho}C^{ab}$. 
 
The Cotton is the  unique tensor which transforms covariantly under conformal diffeomorphism in 3-dimensions, and its determination completely specifies the conformal geometry of a metric (see \cite{Cotton_squashed}). With it, we can form scalars by contraction,  $C^{a_1}_{~a_2}C^{a_2}_{~a_3}\cdots C^{a_n}_{~a_1}$ transforming as $e^{(-6n)\rho}$, since we need to use $n$ metrics to contract all the tensors. Since this is a scalar, we can take the $n$-th  roots, so that we match the conformal factor of $\sqrt{g}$ to make a conformal invariant density. For $n=1$, since the Cotton is traceless, we get just zero. Thus we arrive at the simplest conformally invariant density function: 
$$\sqrt{g} \sqrt{C^{ab}C_{ab}}=\sqrt{g}\, \Xi $$
Where we called $\sqrt{C^{ab}C_{ab}}=:\Xi$, i.e. the \lq{}Cotton norm\rq{}. It is the square root of two terms with three powers of derivative each. However, this term is non-local in a more controlled fashion than non-localities emerging from inverse differential operators. Without the use of a reference density (a background structure), there is no way to formulate a conformally invariant kinetic term, as we now do.

The simplest candidate for a conformal-diffeomorphism invariant Lagrangian is:
\be\label{equ:simple_candidate}\pounds=\int d^3 x\,\sqrt{g}\, \Xi\, (\dot g^{cd}-(\mathcal{L}_\xi g)^{cd}-\rho g^{cd})(\dot g_{cd}-\mathcal{L}_\xi g_{cd}-\rho g_{cd})
  \ee
This Lagrangian is quadratic in the velocities (multiplied by a positive conformal factor), and is invariant wrt the gauge transformations; it thus defines a conformally invariant metric in $\mathcal{Q}$ (see \eqref{equ:conf_supermetric} in appendix \ref{app:PFB})
\be\label{simple_metric} (\mathbf{v, \, w})_g:=\int d^3 x\sqrt{g}\, \Xi\, v_{ab}w^{ab}
\ee
for $\mathbf{v, \, w} \in T_g\mathcal{Q}$. 
The Lagrange multipliers $\xi^a$ and $\rho$ have been inserted in \eqref{equ:simple_candidate} to guarantee invariance under time-dependent gauge transformations, as we explain below. 
 But the same principles could generate:  
  \begin{align}
  \pounds&=\int d^3 x\,\sqrt{g}\Big( (\dot g^{cd}-(\mathcal{L}_\xi g)^{cd}-\rho g^{cd})(\dot g_{cd}-\mathcal{L}_\xi g_{cd}-\rho g_{cd})\sum_{n=0}\alpha_n({C^{a_1}_{~a_2}C^{a_2}_{~a_3}\cdots C^{a_{n+1}}_{~a_1}})^{1/n+1}\nonumber \\  
  &+\sum_m \Lambda_m({C^{a_1}_{~a_2}C^{a_2}_{~a_3}\cdots C^{a_m}_{~a_1}})^{1/m} \Big)+f(CS[g])
\label{equ:general_Lagrangian}  \end{align}
  where $\Lambda_m$ are a generalization of  a cosmological constant,  and $f(CS[g])$ is a general function of the Chern-Simons functional and $\alpha_n$ are arbitrary constants.\footnote{
  We could also add $L(g)=\nabla^2-\frac18R$, the conformal Laplacian,  sandwiched between functions of appropriate homogeneous conformal weight.  That is, for a function $f$ which does not transform under our conformal change, we have $L(e^{4\rho}g)f=e^{-5\rho}L(g)(e^{\rho}f)$, which won\rq{}t be invariant unless $f\mapsto e^{-\rho f}$ and we have another term cancelling the $e^ {5\rho}$ on the lhs of the operator. Similarly, but with more constraints on the coupling, we could add the higher order Paneitz operator \cite{Paneitz}, which also transforms conformally covariantly, provided, again, we wanted to couple functions with another appropriate conformal weight. These can be added and studied in a case by case basis, but they will not be considered here. See \cite{Faci_CG} for a manner of constructing these types of operators.  }

     The least amount of derivatives of the metric we can get (besides zero), with an ultralocal (tensorial) supermetric and no other ingredients,  gives: 
  \be\label{equ:simple_Lagrangian}\pounds=\int d^3 x\,\sqrt{g}\, \Xi\,\left(\Lambda+(\dot g^{cd}-(\mathcal{L}_\xi g)^{cd}-\rho g^{cd})(\dot g_{cd}-\mathcal{L}_\xi g_{cd}-\rho g_{cd})\right)+\alpha CS[g]
  \ee

  \paragraph*{The transformations of $\xi^a$  and  $\,\rho$}
  Under a time dependent  diffeomorphism, $f_t^*g_{ab}$, and under a conformal transformation $e^{\epsilon} g_{ab}$, maintaining invariance would require the transformations (at $t=0$)
   \be\label{equ:Lagrange_transfs}
  \delta \pmb{\xi}= \dot {\pmb{\epsilon}}-[\pmb{\xi},{\pmb{\epsilon}}]\, ~~\mbox{ and}~~\delta \rho= \dot\epsilon-\xi^a\partial_a\epsilon-\epsilon^a\partial_a\rho
   \ee   
where ${\pmb{\epsilon}}:=\epsilon^a\frac{\partial}{\partial x^a}$ is the vector field flow of $f_t$ at $t=0$ and $f_0=$Id (we use boldface to simplify the notation for equations involving the commutator). Thus,   $\xi^a$ and $\rho$ transform in a manner to cancel the extra terms. In the transformations for ${\pmb{\epsilon}}^a$:
\begin{align}
\delta_{\pmb{\epsilon}}{g}_{ab}&=\mathcal{L}_{\pmb{\epsilon}} {g}_{ab}\nonumber\\
\delta_{\pmb{\epsilon}}\dot{g}_{ab}&=\mathcal{L}_{\dot {\pmb{\epsilon}}} g_{ab}+\mathcal{L}_{\pmb{\epsilon}} \dot{g}_{ab}
\label{equ:covariant_conf}
\end{align}
so 
$$\delta_{\pmb{\epsilon}}\mathcal{L}_{\pmb{\xi}} g_{cd}=  \mathcal{L}_{\dot {\pmb{\epsilon}}} g_{cd}- \mathcal{L}_{[{\pmb{\xi}}, {\pmb{\epsilon}}]} g_{cd}+ \mathcal{L}_{{\pmb{\xi}}} \mathcal{L}_{{\pmb{\epsilon}}} g_{cd}=\mathcal{L}_{\dot {\pmb{\epsilon}}} g_{cd}+\mathcal{L}_{{\pmb{\epsilon}}} \mathcal{L}_{{\pmb{\xi}}} g_{cd}$$
The transformations for $\rho$ follow a similar pattern. Putting them together, we get: 
\be\label{equ:cov_transf}
\delta(\dot g_{cd}-(\mathcal{L}_{\pmb{\xi}} g)_{cd}-\rho g_{cd})= \mathcal{L}_{\pmb{\epsilon}} (\dot g_{cd}-(\mathcal{L}_{\pmb{\xi}} g)_{cd}-\rho g_{cd})
\ee
which means that this combination transforms covariantly wrt to conformal diffeomorphisms, a non-trivial result.  The transformation of the shift, $\xi^a$, can be checked to be inherited by a space-time formulation, when identified with components of the spacetime metric under the ADM decomposition.

  But in the present case, in which one is abdicating the space-time view, they must come from a different source, and indeed they do; they arise from the transformation properties of connection-one-forms, in principle fiber bundles (as I  elaborate in appendix \ref{app:horizontal}).  In other words, instead of appealing to a spacetime picture, we merely take the true gauge structure of Riem under conformal diffeomorphisms. It forms a principal fiber bundle \cite{Gauge_riem}, and thus we can use connection 1-forms to covariantize derivatives wrt field-dependent gauge transformations (see figure 6). 
   \begin{figure}[h!]\label{FB}
\center
\includegraphics[width=0.5\textwidth]{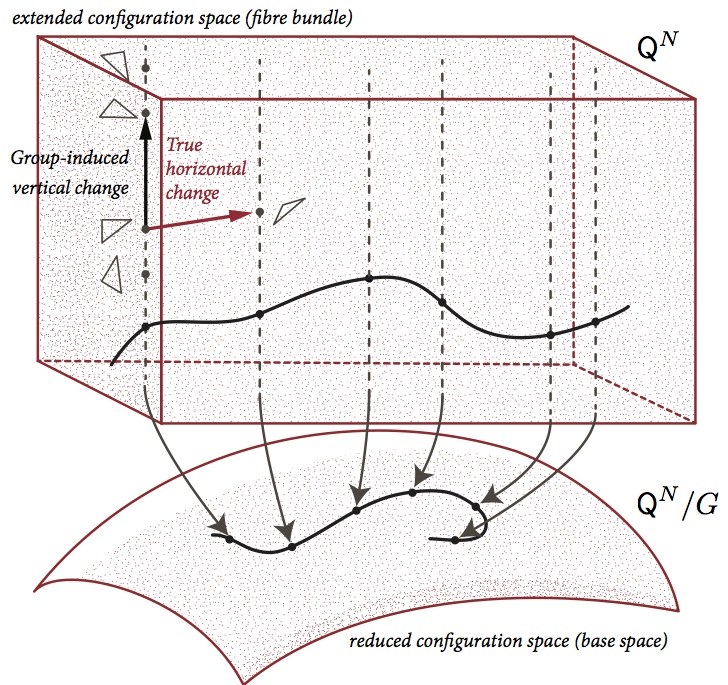}
\caption{An illustration of the physical configuration space in the case of relations between three particles  (triangles). Vertical motion (generated by the symmetries $G$) changes the representation of the triangle in Cartesian space $Q^N$ --- rotations, scalings, and translations ---
while horizontal motion changes the shape of the triangle (e.g., its internal angles). As is standard in the context of principal fiber bundles, horizontality is determined by a choice of connection-form $\varpi$ in configuration space. Figure taken from arxiv \cite{Flavio_tutorial}, with permission.} \end{figure}
     
  In particular,     from \eqref{varpi_tranfo} (derived in all generality in \cite{Aldo_HG}), for spatial diffeomorphisms we would obtain, for the connection 1-form (in field space), $\omega$, under a vertical field-dependent (i.e. time-dependent) transformation:
   \be\delta_{\pmb{\xi}}(\pmb{\omega}(\dot g))=\dot{\pmb{\xi}} -[\pmb{\omega}(\dot g), {\pmb{\xi}}]\ee 
which matches the first of \eqref{equ:Lagrange_transfs}, for $\omega^a(\dot g)$ identified with the shift, $\xi^a$. The connection-form guarantees gauge-covariance. In the case of the group $\mathcal{C}$, it embodies \lq\lq{}parallel transport\rq\rq{} of location along time --- also described as an \lq{}equilocality\rq{} relation \cite{Barbour94} --- and parallel transport of scale along time. 

If we define the connection form by orthogonality to the group orbits with respect to the conformally invariant metric \eqref{simple_metric}, the defining equations for $\xi^a$ and $\rho$ would be:
  \begin{align} \nabla_a\left(\,\Xi\,(\dot{g}^{ab}+\nabla^{(a}\xi^{b)}-\rho g^{ab})\right)&=0\label{equ:xi_eom}\\
\dot {g}_{ab}g^{ab}-\mbox{div}{(\xi)}-3\rho &=0\label{equ:rho_eom}
    \end{align}
 Upon a Legendre transformation, these would equal  the first class constraints of the theory, which form a Lie algebra, as we show below in \eqref{equ:constraints}.

%We can explicitly substitute \eqref{equ:rho_eom} back into the Lagrangian, and using the conformal Killing form, 
 %$$L_\xi g_{ab}=\mathcal{L}_\xi g_{ab}-\frac{2}{3}\mbox{div}(\xi)g_{ab}$$ rewrite \eqref{equ:simple_candidate} in an explicitly conformally invariant manner, without the use of $\rho$:
 %\begin{align}\pounds=&\int d^3 x\,\sqrt{g}d^3x \, \Xi\,\left((\dot g^{cd}-\frac{\dot g}{3}g^{cd})-({L}_\xi g)^{cd}\right)\left((\dot g_{cd}-\frac{\dot g}{3}g_{cd})-({L}_\xi g)_{cd}\right)\nonumber\\
%=&\int d^3 x\,\sqrt{g}d^3x \, \Xi\,\left(\dot g^{cd}-(\mathcal{L}_\xi g)^{cd}\right)_{\st{T}}\left(\dot g_{cd}-(\mathcal{L}_\xi g)_{cd}\right)_{\st{T}} \label{equ:simple_candidate1}\end{align}
  % where $\dot g=\dot g_{ab} g^{ab}$ and \st{T} denotes the traceless projection of the tensor. 
   
 \paragraph*{Hamiltonian version}
  Using \eqref{equ:xi_eom} to get rid of extra terms that would arise from taking the variational derivative of the connection form wrt $\dot g_{ab}$, we have
 \be
 \label{equ:momenta}
\frac{\delta \mathcal{L}}{\delta \dot g_{cd}}=: \pi^ {cd}\hat{=}\sqrt{g} \, \Xi\,(\dot g^{cd}-(\mathcal{L}_\xi g)^{cd}-\rho g^{cd})
 \ee 
 where the hat means we have used \eqref{equ:xi_eom}. 
 
The Hamiltonian form of \eqref{equ:simple_Lagrangian} is easily worked out to be: 
  \be\label{equ:simple_Hamiltonian}\mathcal{H}=\int d^3 x\,\left(\frac{\pi^{ab}\pi_{ab}}{4\sqrt{g}\, \Xi\,}-g_{ab}\mathcal{L}_\xi\pi^ {ab}-\rho g_{ab}\pi^ {ab} -\sqrt{g}\, \Xi\,\,\Lambda\right) -\alpha CS[g]
  \ee
  The transformations \eqref{equ:Lagrange_transfs} guarantee that the action:
  $$S=\int \pi^{ab}\dot g_{ab}-\mathcal{H}$$ 
  is invariant wrt to generalized field-space dependent conformal diffeomorphisms. 
  
If we furthermore want the action to be invariant wrt the choice of field-space connection 1-form, we obtain the usual form of the constraints. As is the case in general relativity, their content is, as expected, that the gravitational momentum is transverse and traceless,\footnote{An interesting project is to define the theory wrt a particular connection form, i.e. with a particular horizontal lift. One would have a non-local action which is diffeomorphism invariant, but no associated free Lagrange multiplier. This has not been explored.}
 \be\label{equ:constraints}\mbox{ $D_b(x):=\nabla_a\pi^{a}_{~b}(x)=0$ ~ and ~ $D(x):=\pi^{ab}g_{ab}(x)=0$}
 \ee      
 The commutator relations, with smearings $\xi^a$ and $\rho$ are:
 \be\label{equ:constraints_com}\mbox{ $\{D_b(\xi_1^b),\, D_b(\xi_2^b)\}=D_b([\xi_1,\xi_2]^b)$\,  , ~\, $\{D_b(\xi^b),\, D(\rho)\}=D(\xi^a\partial_a\rho)$\, ~ and ~\,$ \{D(\rho_1),\, D(\rho_2)\}=0$}
 \ee 
The equations of motion of the metric and the momenta are:
\begin{align}
\dot{g}_{ab}&=\frac{\pi_{ab}}{2\sqrt{g}\, \Xi}-\mathcal{L}_\xi g_ {ab}-\rho g_{ab}\label{equ:eom_g}\\
\dot{\pi}^{ab}&=\frac{1}{2\sqrt{g}\, \Xi}\left(\pi^{ac}\pi_{c}^b-\frac{\pi^{ij}\pi_{ij}}{ \Xi^2}(C^{ac}C_c^b+C^{cd}\diby{C_{cd}}{g_{ab}})\right)\nonumber\\
&-\Lambda\sqrt{g}\left(\frac{\Xi}{2} g^{ab}+\frac{1}{\Xi} C^{cd}\diby{C_{cd}}{g_{ab}}\right)-\alpha C^{ab}-\mathcal{L}_\xi \pi^{ab}-\rho \pi^{ab}
\label{equ:eom_pi}
\end{align}
The functional derivative of the Cotton tensor is rather involved, and we thus leave its form implicit. 
 \subsection{Coupling to matter}\label{sec:matter}
 The coupling of matter is usually determined by minimal coupling in general relativity. It  is an important issue to settle under the new management of different symmetry principles. However, it is too complicated an issue to be discussed to completion here. I leave a devoted study for the near future.

  The solution employed by York  et al. \cite{York3} and Isenberg et al \cite{IsenbergNester2} for the conformal weight of extra matter fields, was to make them zero. That is, the extra fields are then only carried along by the transformations of the metric. This automatically makes the usual Yang-Mills Lagrangian conformally covariant at least.

    %   If we want to put everything on the same footing as the metric variables,  we should have  $A_a\rightarrow e^{2\phi}A_a$, so that the kinetic term stays invariant, and thus also gets multiplied by the factor $\sqrt{g} \, \Xi\,$.  Similarly to \eqref{equ:covariant_conf},  we have, e.g.:
% $$(\dot A_{a}-(\mathcal{L}_\xi A)_{a}-\frac{\rho}{2} A_{a})\rightarrow e^{2\phi}(\dot A_{a}-(\mathcal{L}_\xi A)_{a}-\frac{\rho}{2}A_{a}) $$ with the same $\rho$ and $\xi$, and same transformation properties, as in \eqref{equ:Lagrange_transfs}.   In the momentum representation, $E^a:=p_{\tiny A}^a$, this means that: $E^a\rightarrow e^ {-2\phi} E^a$, which is also the induced transformation for the vector field $A^a$.   {Another alternative is that we could have the kinetic term of the field have the opposite transformations as the volume density. In that case, we would have $A_a\rightarrow e^{-\phi}A_a$, and thus $A^a\rightarrow e^{-5\phi}A^a$, while  $E^a\rightarrow e^ {\phi}E^a$, which is a less symemtrical rule than the previous transformation. } With either of these scalings, we need to find potential terms that are spatially conformally invariant, which might be a non-trivial exercise. 
 \paragraph*{Coupling of massless vector potentials.}
  Under a conformal transformation $g_{ab}\rightarrow e^{4\phi}g_{ab}$, terms that depend on the derivative of $\phi$ would appear in the Christoffel symbols, and only highly special combinations of these can get rid of such derivative terms (i.e. the Cotton tensor).   
  Thus for (possibly Lie-algebra valued) one forms $A_a$, we need to have anti-symmetry in spatial derivatives, as in $\nabla_{[a}A_{b]}\nabla^{[a}A^{b]}$, so that the corresponding term can be conformally covariant. That is because anti-symmetrized covariant derivatives of vector fields don\rq{}t register different Christoffel symbols (for Levi-Civita connections), $$\nabla_{[a} A_{b]} =\partial_{[a} A_{b]}+\Gamma^c_{[a b]} A_c=\partial_{[a} A_{b]}:=B_{ab}$$
   In other words, no symmetrized spatial derivatives of spin-1 forms can appear, since otherwise there would be no way to make the coupling conformally covariant without the introduction of a further Weyl-connection \cite{Weyl}, and this we would like to avoid (see \cite{SD_coupling}).\footnote{ Weyl\rq{}s notion of conformal potential to correct for derivatives of the conformal factor would change the analysis through the introduction of another field -- the one Weyl tried to identify with the electromagnetic potential. The introduction of this field would however also allow for a different set of vacuum Lagrangians. }

 If there are no terms of the form $A_a g^{ab} A_b$, then the potential for the field can only appear with anti-symmetric spatial derivatives described above; the symmetrized part is not dynamical. There must thus be  a redundancy of description. This is just the symmetry $A_a\rightarrow A_a+\partial_a\psi$, in the case of $U(1)$ for some scalar function $\psi$. 
  By the usual arguments of gauge theory, this redundancy elicits the introduction of a spatial Gauss constraint. Indeed, as elaborated in appendix \ref{app:horizontal}), we would replace time derivatives by \lq{}gauge-corrected\rq{}, or horizontal derivatives, 
  $$\dot A_a\rightarrow \dot A_a -\partial_a \omega $$
 where,  under a vertical field-dependent (i.e. time-dependent)  $U(1)$ transformation $\psi$,  the field-space connection one-form  $\omega$ transforms, from the equivalent of \eqref{varpi_tranfo}, as $\omega\rightarrow \dot \psi$, i.e. it transforms just as $A_0$ would in a covariant picture. Here, again,  instead of appealing to a spacetime picture, we merely take the true gauge structure of the field space of spatial vector potentials, $\mathcal{A}$, under $U(1)$, finding covariant structures as in the spacetime picture.

  Let the conjugate density to $A_a$ be $E^b$, for which we define the undensitized $\bar E^b\sqrt{g}=E^b$. These are the fields that are required to be conformally invariant.  In this case the Gauss constraint doesn\rq{}t require a compensating field for the conformal transformation since
   $$\sqrt{g} \nabla_a \bar E^a=\sqrt{g}\frac{1}{ \sqrt g}\partial_a (\sqrt g\bar E^a)=\partial_a E^a
  $$
Following the usual recipe, the conformally invariant Hamiltonian for this field up to quadratic order becomes:\footnote{Were we to  include internal indices, we would merely have to also include the Killing inner product for the Lie algebra at hand in the Hamiltonian, and $\nabla_{[c}A_{d]} \rightarrow \nabla_{[a}A^I_{b]}+\alpha f^I_{~JK}A^J_a\,A^K_b$ which still transforms in the same way.}:   
\begin{align}   \label{equ:vector_Hamiltonian}\mathcal{H}_{\mbox{\tiny vec}}&=\int d^3 x\,\Big(\frac{\alpha_{\mbox{\tiny v}}E^a\,g_{ab} E^b+\beta_{\mbox{\tiny v}} g B_{ab}\,g^{ac}g^{bd}\,B_{cd}}{\sqrt{g}\, {\Xi}^{1/3}}-A_a\mathcal{L}_{\xi}E^a+\lambda\partial_aE^a\Big) 
  \end{align}
where $\alpha_{\mbox{\tiny v}}, \,\beta_{\mbox{\tiny v}}$ are coupling constants, the scaled permittivity and the magnetic permeability, respectively, i.e. $[\beta_{\mbox{\tiny v}}]\sim [\mu_0]/L$ and  $[\alpha_{\mbox{\tiny v}}]\sim [\epsilon_0]/L$ (note that $[\Xi^{1/3}]=L^{-1}$). Note that, since only the metric transforms under conformal transformations, both terms $E^a\,g_{ab} E^b$ and $g\nabla_{[a}A_{b]}\,g^{ac}g^{bd}\,\nabla_{[c}A_{d]}$ have the same conformal weight, $e^{4\phi}$, and thus require the same compensation by the denominator.

 Besides the presence of $U(1)$ symmetry, these are other \lq\lq{}coincidences\rq\rq{} between standard relativistic electromagnetism and this, spatial conformal coupling to vector field Hamiltonians.  In other words,  conformal covariance and requiring the two terms in the Hamiltonian to have the same weight requires some sort of relativity principle --- we could only match even powers of $E$ with even powers of the anti-symmetrized $\nabla A$. 

Apart from the presence of $\Xi$, the equations of motion for $E$ and $A$ assume  the standard form:
\begin{align}   \label{equ:eom_EM}
\dot E_a &= \beta_{\mbox{\tiny v}}\sqrt{g}\, \nabla^b\left(\frac{\nabla_{[a}A_{b]}}{\Xi ^{1/3}}\right) -\mathcal{L}_\xi E^a
\\ 
\dot A_a &=  \frac{\alpha_{\mbox{\tiny v}}}{\sqrt{g}\,\Xi ^{1/3}} E_a+\partial_a\lambda+\mathcal{L}_\xi A_a
  \end{align}
From these two equations we can get back to the Lagrangian equations of motion (setting $\xi^a=0=\lambda$):
\begin{align} \label{full_EM}
\ddot A_a =   \frac{\alpha_{\mbox{\tiny v}}\beta_{\mbox{\tiny v}}}{\Xi ^{2/3}}\left(\nabla^b\nabla_{[a}A_{b]}-\frac13 \nabla_{[a}A_{b]}\nabla^b \ln \Xi\right) -\dot A_a\left(\frac{\dot g}{2}+\frac{1}{3 }\frac{d}{dt} \ln \Xi\right)
  \end{align}
  where $\dot g=g^{ab}\dot g_{ab}$.
In the adiabatic limit for the scale part of the geometry ($\dot g=0$, which is usually part of the gauge for the metric dofs) rewriting the wave-equation in terms of the vector potential, assuming that ${\nabla_a \ln \Xi}$ is small when compared to the gradients of the magnetic field, schematically $|\nabla_a \ln \Xi| $ and in the divergence-free gauge ($\nabla^a A_a=0$), we obtain:
\be\label{EM_wave}  \ddot A_a=\frac{\alpha_{\mbox{\tiny v}}\beta_{\mbox{\tiny v}}}{\Xi ^{2/3}}(\nabla^2 A_a-R^b_a A_b) \ee
which is the standard  form of the wave-equation for the vector potential in curved space-time (for $c=1$)
$$ \square A^{\alpha}-R^\alpha_\beta A_\alpha=0
$$
(for the space-time Ricci tensor and the covariant D\rq{}Alembertian) when $A_0=0$ and the extrinsic curvature is transverse and traceless (by the contracted Gauss-Codazzi equation, this implies that $R^0_a=0$), where 
$$\left[\frac{\alpha_{\mbox{\tiny v}}\beta_{\mbox{\tiny v}}}{\Xi ^{2/3}}\right]\sim [\mu_0\epsilon_0]\sim[c^2]$$

  One possible line of investigation regards the degree to which local inhomogeneities  in the gravitational field (parametrized by ${\nabla_a \ln \Xi}$) influence the standard curved spacetime electrodynamics equation. Note that, having been gauge-fixed, the equations of motion \eqref{EM_wave} need no longer be conformally invariant.

\paragraph*{Coupling of massless scalar fields.}
Accordingly, if we want to use the same principles for a scalar field Hamiltonian, for a simple kinetic term of the form $p_\varphi^2$,  a potential term of the form $\partial_a\varphi\, g^{ab}\,\partial_b\varphi$ and a mass term of the form $m^2\varphi^2$,  we obtain the anisotropic Hamiltonian:
%  \begin{align}    \label{equ:scalar_Hamiltonian1}\mathcal{H}^{(1)}_{\mbox{\tiny scl}}=\int d^3 x\,\Big(\frac{p^2_\varphi+\alpha_{\mbox{\tiny s}} g\,(\partial_a\varphi\, g^{ab}\,\partial_b\varphi)^{3}}{\sqrt{g}\, \Xi\,}-\varphi\mathcal{L}_{\xi} p_\varphi     -\sqrt{g}\, \Xi\,\,m^2\varphi^2\Big)   \end{align}
%   We could demand that the scalar gradient term come also with the second power, but this would either break conformal invariance, setting a fixed scale to the system, or come as a sub-leading term in an inverse $C^{ab}C_{ab}$ expansion:   
 \begin{align}   \label{equ:scalar_Hamiltonian2}\mathcal{H}_{\mbox{\tiny scl}}=\int d^3 x\,\Big(\frac12\frac{p^2_\varphi}{\sqrt{g}\, \Xi\,}-\varphi\mathcal{L}_{\xi} p_\varphi
    -\frac12\sqrt{g}\left(\, \Xi\,\,m^2\varphi^2-{\beta_{\mbox{\tiny s}}}\, \Xi^{1/3}\,\partial_a\varphi\, g^{ab}\,\partial_b\varphi\right)\Big) 
  \end{align}  Given the possible conformal weights, we need to classify the types of matter potentials and kinetic terms according to the analogous equations \eqref{equ:counting_tensor} and \eqref{equ:counting_conf}. In an effective field theory framework, we need to add all the terms compatible with the symmetries, which would include more complicated derivative couplings. %\footnote{In the asymptotic safety scenario, it is shown that more general non-derivative couplings are generated through the renormalization group flow of minimal coupling terms with the Einstein-Hilbert action \cite{Astrid_matter}. }    
   These are all interesting questions, and there seems to be a lot of room for phenomenological study of this dynamical hierarchy, which is all  left for the future. 
  
  The equations of motion are: 
  \begin{align}   \label{equ:eom_phi}
\dot \varphi &= \frac{p_\varphi}{\Xi\, \sqrt{g}}-\mathcal{L}_\xi \varphi
\\ 
\dot p_\varphi &= \left( \sqrt{g}\, \Xi\,\,m^2\varphi-{\beta_{\mbox{\tiny s}}}\,\partial_b\left( \sqrt{g}\,\, \Xi^{1/3}\,g^{ab}\,\partial_a\varphi\right)\right)-\mathcal{L}_\xi p_\varphi
  \end{align}
Note that the equations of motion are explicitly conformally invariant, and that the leading term in $\dot p_\varphi$ for $\Xi<<1$ is the kinetic one, not the mass one. For $\xi=m=0$, and $p_\varphi$ a constant in time,  we obtain the elliptic, conformally invariant equation: 
\be\label{New_LFE} \partial_b\left( \sqrt{g}\,\, \Xi^{1/3}\,g^{ab}\,\partial_a\varphi\right)=0\Rightarrow  \frac13\partial_b\ln (\, \Xi)\,g^{ab}\,\partial_a\varphi+\nabla^2\varphi=0
\ee
 where $ \frac13\partial_b\ln (\, \Xi)\,g^{ab}\,\partial_a\varphi+\nabla^2\varphi$ is only conformally covariant.  As in the electromagnetism case, for correspondence with GR we will work in approximations where $\partial_b\ln (\, \Xi)$ is small, which fixes gauge transformations to the extent that $\partial_b\rho<<\rho$.\footnote{Which is not a conformally invariant condition I.e. it only makes sense in some gauge, analogously to statements that, without extra symmetry assumptions, $R_{abcd}$ is small (it is also only a diffeomorphism \textit{covariant} quantity.)} 
  \paragraph*{A simple approximate classical cosmological solution: Bianchi IX.}  
    
 Spherical symmetry implies that the Cotton tensor vanishes, and so any expansion around such a metric for the Hamiltonian \eqref{equ:simple_Hamiltonian} is subtle.   Nonetheless, there are cosmological models, mostly based on Bianchi IX, which are anisotropic and carry shape degrees of freedom (see \cite{Through_BB} for a discussion of these models in the context of shape dynamics). The Bianchi IX spatial metric is of the (abstract-index) form:
 \be\label{BianchiIX}
g_{ab}^{\mbox{\tiny IX}}=a_1^2\sigma_1\otimes \sigma_1+ a_2^2\sigma_2\otimes \sigma_2 +a_3^2\sigma_3\otimes \sigma_3
 \ee
 where $d\sigma_i=\frac{1}{2}\epsilon_{ijk}\sigma_j\wedge \sigma _k$ are left-invariant one-forms, defined on the  3-surface $M$ (here considered to be 3-spheres). 
In principle the coefficients $a_i$ can have arbitrary dependence on time, however, here I will limit myself to the regime where this dependence is very slow. 
  Since we have a conformally invariant model, the parametrization \eqref{BianchiIX} just forms \lq\lq{}a section\rq\rq{} for our metric; we can multiply this metric arbitrarily by a factor $\phi$. We will fix this factor as follows.  
 
 Let us take such a history of 3-metrics, $g_{ab}(t)$ in a gravitational adiabatic limit, so that \footnote{Here, gravitationally adiabatic means that, if coupled to another source, e.g. electromagnetism, as above, the rate of change $\dot g_{ab}$ is small when compared with $E^a$. Of course, this includes the static case, $\dot g_{ab}=0$. } $\order{\pi^{ab}}\sim\epsilon$, and choose $\alpha=\rho=\xi=\Lambda=0$. 
 It is easy to see that this is an approximate (up to order $\epsilon$) solution for the equations of motion \eqref{equ:eom_g}-\eqref{equ:eom_pi}: 
\begin{align}
\dot{g}_{ab}&=\frac{\pi_{ab}}{2\sqrt{g}\, \Xi}\sim \order{1}\\
\dot{\pi}^{ab}&=\frac{1}{2\sqrt{g}\, \Xi}\left(\pi^{ac}\pi_{c}^b-\frac{\pi^{ij}\pi_{ij}}{ \Xi^2}(C^{ac}C_c^b+C^{cd}\diby{C_{cd}}{g_{ab}})\right)\sim  \order{\epsilon}
\end{align}
as long as the conformal geometry stays approximately flat, to the extent that $\mathcal{O}({\Xi})\sim \order{\epsilon}$. 

For traceless momenta and divergence-free $\xi$, setting  $\rho=0$ means that the trace $\dot g=\dot g_{ab} g^{ab}$ is constant.   We can set the conformal factor to be determined by the local volume form of the unit round 3-sphere. % % eventually we want the curve of geometries to cross that point (see below),\footnote{The limit is indeed very delicate, since there the Cotton tensor vanishes. This would also have extreme implications for the quantum theory, together with our \lq\lq{}past hypothesis\rq\rq{} (see section \ref{sec:past}).} $g_{ab}(0)=\Omega_o$ (see section \ref{sec:past} above), and thus $\det g=\det \Omega_o$. 
{A priori, solutions of the equations of motion merely give us a field-history such as  $g_{ab}(t)$. To recover a spacetime, we need to establish how the instantaneous configurations connect along time. The connection-form $\varpi$, discussed in appendix \ref{app:PFB} (see \eqref{varpi_tranfo}) provides the ``shift'' part of the metric. }

From here on we assume that $\dot g=\dot g_{ab} g^{ab}=0$.  Then, assuming ${\nabla_a \ln \Xi}$ is small when compared to the gradients of the magnetic field, \eqref{full_EM} becomes  \eqref{EM_wave}, and  we can use the propagation of vector fields to \lq\lq{}reconstruct a space-time metric\rq\rq{}.  That is, using  the characteristics of the electromagnetic wave-equation \eqref{EM_wave} to build a light-cone, we can also reconstruct  a ``duration'', or lapse. We thus  have a 4-metric approximately of the standard Bianchi IX form (with zero shift, i.e under horizontal lifts by some connection-form):
\be\label{4-metric}
ds^2=-N^2dt^2+ g_{ab}^{\mbox{\tiny IX}}
\ee
where $N^2= \frac{\alpha_{\mbox{\tiny v}}\beta_{\mbox{\tiny v}}}{\Xi ^{2/3}}$ is of the same dimension as $c^2$. 
Of course, away from the adiabatic limit of the geometry,  we would not have such a simple solution at hand, and with inhomogeneities in the Cotton norm (i.e. when $|\partial \Xi| >>|\Xi|$) we couldn\rq{}t use \eqref{EM_wave} and the simple form for reconstructing spacetime would no longer be available. On the up-side, this result gives some hope in modifying quantum cosmology in a regime-dependent manner, where inhomogeneities in space and time affect the speed of light for example.  

  \paragraph*{Approximating Schwarzschild.}  

There are further challenges to this theory when it comes to obtaining classical solutions which rely on spherical symmetry, such as an analogue of a Schwarzschild solution to model the solar system. Here the solution is to slightly deform the spherical metric, given by $a_1=a_2=a_3$ in \eqref{BianchiIX}, to $a_1\rightarrow \kappa+\epsilon a_i$.  In that case,  we can arrive at an estimate for  $\order{\Xi}$ from its conformal weight. Namely, under a conformal transformation $\rho g_{ab}$ the Cotton norm itself would obtain a term $\rho^{-3}$ (which cancels the similar term coming from the determinant of the metric). Since $\epsilon$ is deemed to be diagonal in the metric, and  it is taken as an infinitesimal transformation, we obtain that  $\Xi \sim \epsilon$. According to the previous section,  we can solve the equations of motion for the metric as long as $\pi_{ab}\sim \epsilon$ (for $\Lambda=\alpha=\xi=0$). 

We will use the scalar field to construct the lapse in this approach. We will choose the simplest possible such construction under spherical symmetry, with $\dot\varphi=N$, and $\varphi(r,t)=\varphi(r)t$. Since we are assuming that the metric degrees of freedom are also static, using \eqref{equ:eom_phi} this  means that $\dot p_\varphi=0$.

Spherical symmetry for the background implies that that background is conformally flat.\footnote{This is true even if the manifold is spatially closed, since a Sobolov conformal factor can be used to decompactify it.} We thereby set $g_{ab}=\Omega^4\delta_{ab}$ where $\delta_{ab}$ is the standard flat metric in spherical coordinates, $\delta_{ab}=\left(dr^2+r^2(d\theta^2+\sin^2(\theta)d\varphi^2)\right)$. For the conformal factor, we assume it is analytic, and well-behaved at infinity, thus: $\Omega=\sum_{i=0} m_i/r^i$. Here the dimensionful constants $m^i$ are setting the relevant scales to the system.  Now, setting $\xi^ a=0$, since $\dot p_\varphi=0$,  and in the approximations where $\partial_b \ln \Xi<<1$ in a particular gauge, we can use \eqref{New_LFE}, obtaining: 
$$\partial^ 2\varphi+2\partial_a\Omega\,  \delta^{ab}\partial_b\varphi\approx 0$$
The requirement $\partial_b \ln \Xi<<1$ also implies a domain of validity for $\Omega$.  Setting $m_1=m$ and $m_i=0, \forall i>1$, this implies that $r>>m$.  It can be checked that, for large enough distances, the solution to this equation is $\varphi=1-\frac{2b}{m+2r}+\order{r^{-2}}$, with $b$ another integration constant of the same dimensions as $m$ (and $r$), which we set to $m$ as well. This \lq{}coincidence\rq{}  is not explained here. We can then rewrite to this order $\varphi=N=\left(\frac{1-\frac{m}{2r}}{1+\frac{m}{2r}}\right)$, obtaining  
\be \label{equ:SD_new}
ds^2= -\left(\frac{1-\frac{m}{2r}}{1+\frac{m}{2r}}\right)^2dt^2+ (1+\frac{m}{2r})^4\left(dr^2+r^2(d\theta^2+\sin^2(\theta)d\varphi^2)\right)
\ee
This is the Schwarzschild solution in isotropic coordinates. Although this is encouraging, it does not yet have the same significance as the Schwarzschild solutions, as we have not shown an equivalent of the Birkhoff theorem (as has been done for shape dynamics \cite{SD_birkhoff}), and we cannot know if it is the metric forming from collapse, and it also should not be valid for small radii. Nonetheless, it seems that in the approximations we are enforcing, one can obtain  standard phenomena related to GR in the solar system. 
   
Deviations from spherical symmetry are required so that the equations of motion are well-defined, but we found approximations (in both field values and spatial domains) in which the specific form of these deviations would only show up on;y at higher order of the equations of motion. 
\section{The path integral}\label{app:conformal_action}

To concretely discuss this path integral, one would have to specify the measure of integration specifically, and describe a regularization procedure. We  won\rq{}t have much to say about either regularization or integration measures at this point. The only point which we will dwell on is that of gauge-fixing Sec. \ref{sec:gf}, BRST transformations Sec. \ref{sec:BRST}, and invariance of the wave-function Sec. \ref{sec:inv}, since they illustrate the ease with which one can treat the spatial conformal diffeomorphisms, as opposed to the spacetime diffeomorphisms.

We  want to define a wave-function anchored on some preferred boundary conditions on physical configuration space. That is, given a preferred physical configuration $[g_o]$, schematically (see \eqref{wave-function_scheme} below for more details)
\be  \Psi([g]):=\int  \mathcal{D}[\gamma] \exp{i S([\gamma([g_o], [g])])/\hbar}
\ee
I.e. we perform a path integral in reduced configuration space connecting this \lq{}anchor\rq{} to the given $[g]$. We will discuss the role and definition of $[g_o]$ in section \ref{sec:anchor}, below.

\subsection{Preferred strata and variational principles for boundary conditions.}\label{sec:anchor}

In Hartle-Hawking, there is some subtlety on how one goes about this integration,\footnote{A 3+1 decomposition is utilized for concrete calculations, and the meaning of the Euclidean path integral becomes more subtle.} but $[g_o]$ is generally regarded as related to the completely degenerate metric, $g_{ab}=0$. In our case, things are different. 

In general relativity, there is no intrinsic action of the local symmetries on configuration space (i.e. the action of the refoliations don\rq{}t project down to configuration space, as illustrated by equation \eqref{refol_sym} and figure 6). Thus, specifying conditions on the quotient space of configurations doesn\rq{}t have any meaning. Here, we have the opposite state of affairs, and the stratification above has  salient significance.  

In particular, one can single out configurations with the highest possible dimension of the stabilizer subgroup,  the most homogeneous configurations, as we saw in section \ref{sec:past}. Such a prescription would have no gauge-invariant meaning in a generally covariant theory. Let us elaborate.
The set defining the corners of physical configuration space is:
\be\label{anchor_strata}\Phi_o=\{ \mathcal{O}_g\subset \mathcal{Q} ~|~\mbox{I}_g\subset \mathcal{C}~~ \mbox{has maximum dimension} \}\ee
 Such a set is composed of the least dimensional, and simplest --- in the sense that they correspond to the most homogeneous configurations --- corners of reduced configuration space.  It is these preferred singular points of configuration space that I will define as an origin of the transition amplitude, or the \lq\lq{}anchor\rq\rq{} of the path integral ---- as is done in Hartle-Hawking \cite{Hartle_Hawking}, Vilenkin \cite{Vilenkin}, and Linde \cite{Linde} --- below.  Unlike Hartle-Hawking, however, such boundary  conditions on $\mathcal{Q}/\mathcal{C}$ are physical, and extremize a given gauge-invariant quantity; for instance, the dimension of the strata (or the dimension of the isotropy group).

    Depending on the symmetries acting of configuration space, and on the topology of $M$, one can have different such preferred configurations. %For diffeomorphisms, i.e. for the quotient $\mathcal{Q}/\mbox{Diff}(M)$,  one can have (the equivalence class of) the completely degenerate metric and also fields which have zero value. This would be the \lq\lq{} nothing\rq\rq{}  configuration, there is no distance between any point of $M$. 
    For the case at hand -- in which we have both scale and diffeomorphism symmetry and  $M=S^3$--- there exists a \emph{unique} such preferred point. It is easy to show that in this case $\Phi_o=\{[\Omega_o]\}$ where $\Omega_o$ is the round sphere metric (see e.g. \cite{Giulini_geometrodynamics}),\footnote{For this analysis I assumed that no degenerate metrics are allowed. If they are allowed, then the analysis differs. The completely degenerate metric would be a natural candidate in this extended configuration space, as it possesses the full group of diffeomorphisms as a stabilizer. However, it is not clear how to connect the completely degenerate metric  to the rest of reduced configuration space.  } as it has the maximum number of conformal Killing vector fields.    Differently than before, the boundary conditions are given by an extremum principle, which works at a topological level. Are there other sorts of gauge-invariant functionals we could have chosen to minimize? Are there other variational criteria of boundary conditions for our wave-function? 
    
    Indeed, one could also obtain the same implementation of a variational principle for the preferred boundary conditions using any conformally invariant functional. However, as we saw in section \ref{sec:form_Lagrangian}, these are not so easily constructed. Let us look at three known examples.
    
      We start with  the Yamabe functional \cite{Yamabe}: $Y[g]\rightarrow \mathbb {R}$, given by 
\be\label{Yamabe} Y[g]:=\inf_{\theta}\frac{\int d^3x\, \sqrt{g}\left((\nabla\theta)^2+\frac18 R\theta^2\right)}{\int d^3x \sqrt{g} \theta^6}^{1/3}
\ee
which is invariant under the simultaneous transformations $g_{ab}\rightarrow e^{4\alpha}g_{ab}$, $\theta\rightarrow \theta/\alpha$ and another is the Chern-Simons functional given in \eqref{equ:CS}.  And yet another is the following integral:
\be
\int d^3\, \sqrt{g} \sum_m \beta_m({C^{a_1}_{~a_2}C^{a_2}_{~a_3}\cdots C^{a_m}_{~a_1}})^{1/m}
\ee
where $\beta_m$ are constants. In all of these examples, the conformal equivalence class of the round sphere extremizes the functional. In the first two cases, they are the unique extrema.

  Let us quickly show that this is the case for the Yamabe functional (see also \cite{Yamabe_Murchadha}). Given the extremizing scale $u$, and $\bar g_{ab}=u^ 4g_{ab}$,  in \eqref{Yamabe}, the functional derivative of the Yamabe constat  is given by:
  $$\delta_h Y[g]=\int d^3x\, \sqrt{g} \left(-\frac18 u^2R^{ab}(\bar g)+\frac13 Y[g]u^6g^{ab}\right)h_{ab}$$
  and thus extremization yields
   $$-3u^2R_{ab}(\bar g)+Y[g]u^6g_{ab}=-R_{ab}(\bar g)+Y[g]\bar g_{ab}=0$$
  Since $Y[g]$ is a constant, the only solution is that $\bar g_{ab}$ has constant Ricci curvature; i.e. it is the round sphere. Indeed, the Yamabe functional was the first tentative analogue for shape dynamics of the \lq\lq{}complexity functional\rq\rq{} used in the particle model to obtain the arrow of time \cite{Barbour_arrow}.

    For the Chern-Simons functional, one must merely know that its functional derivative wrt the metric gives the Cotton tensor, which vanishes only for conformally flat metrics. These all give converging arguments for  the round 3-sphere as  the variationally selected boundary conditions for the wave-function of the Universe. 
  
The conclusions up to now are valid also for shape dynamics. In fact, they can be used to quantize the theory as is advocated in \cite{QG_deco}. Now, we will move on to theories based on local Lagrangians.\footnote{A different construction principle exists for shape dynamics, not reliant on locality, but on the existence of mutually gauge-fixing symmetries \cite{SD_construction}. That result, when taken in light of \cite{Vasu_rigidity}, implies uniqueness of shape dynamics up to theories of second order in gravitational momenta. }

\subsection{Gauge fixation}\label{sec:gf}
For completion of our illustration of the formalism under these new symmetry principles,  will now see how much simpler the formal (and abstract) study of symmetries under the path integral quantization may become.  

The action of the present simple model is given through \eqref{equ:simple_Hamiltonian} simply as 
\be\label{classical_action}S=\int dt\left(\int d^3 x\,  \dot g_{ab}\pi^{ab}-\mathcal{H}\right)\ee

The local gauge transformations, under the parameters $\epsilon^a(x)$ and $\epsilon(x)$  are given by (from \eqref{equ:Lagrange_transfs}-\eqref{equ:covariant_conf}):
\begin{align}
\delta{g}_{ab}(x)&=\mathcal{L}_{\pmb{\epsilon}} {g}_{ab}(x)+\epsilon g_{ab}(x)\nonumber\\
\delta{\pi}^{ab}&=\mathcal{L}_{\pmb{\epsilon}} {\pi}_{ab}(x)-\epsilon \pi_{ab}(x)\nonumber\\
  \delta \pmb{\xi}&= \dot {\pmb{\epsilon}}-[\pmb{\xi},\pmb{\epsilon}]\nonumber\\
 \delta \rho &= \dot\epsilon-\xi^a\partial_a\epsilon-\epsilon^a\partial_a\rho
\label{equ:all_transf}
\end{align}

In order to path integrate the exponential of $i S$, we must select a unique representative among each field history related by iteration of the infinitesimal transformations \eqref{equ:all_transf}. The simplest possible gauge condition one can choose is 
\begin{eqnarray}
\xi^a&=0 \label{shift_gf}\\
\rho &=0 \label{conf_gf}
\end{eqnarray}
These could be implemented by adding the Gaussian term 
\be\label{S_gf} S_{\mbox{\tiny g.f.}}=\frac{1}{\sigma}\int dt\int d^ 3 x \sqrt{g} \left(\rho^ 2+ \xi^a g_{ab} \xi^b\right)
\ee
and taking the limit $\sigma\rightarrow 0$.

   Note that in GR, it is not possible to choose the gauge for the lapse as $N=0$ (or in our notation of equation \eqref{refol_sym}, $\lambda^\perp=0$), since this would also \lq\lq{}kill the dynamics\rq\rq{} (again, here a problem of mixing between dynamics and gauge freedom).\footnote{It is possible nonetheless, to choose such a gauge for the perturbation fields in the background field formalism \cite{Barvinsky_hor}.}

In whichever manner we rebuild spacetime, condition \eqref{shift_gf} tells us that the line joining two points with the same spatial coordinates lying on two neighboring surfaces will be normal to the first surface. This condition, together with \eqref{equ:constraints}, \eqref{equ:eom_g} and condition \eqref{conf_gf}, tell us that the metric won\rq{}t change its local volume form along time. In fact, these conditions are equivalent to having the choice of curve in $\mathcal{Q}$ representing the curve in $\mathcal{Q}/\mathcal{C}$ be given by a horizontal lift, according to a connection in the principal fiber bundle, as described briefly in appendix \ref{app:horizontal} (see figure 4). 
\begin{figure}[h]\label{fig:PFB}
\begin{center}
\includegraphics[width=.6\textwidth]{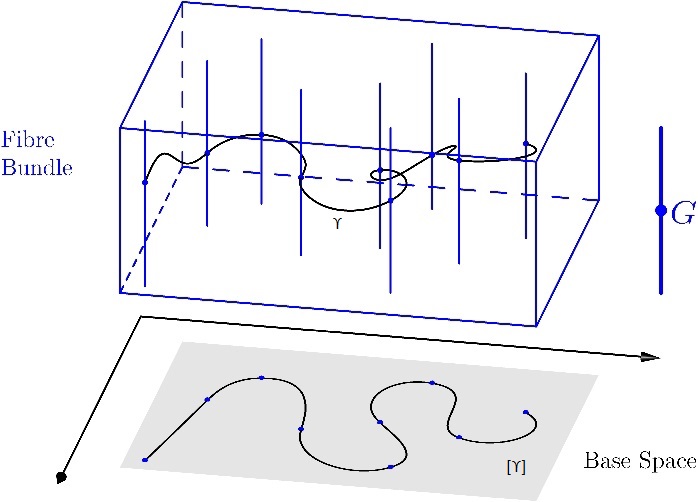}
\caption{The principal fiber bundle picture with group $G$, showing a horizontally lifted curve from the base space (conformal superspace) into configuration space ($\mathcal{Q}$). (this picture ignores the stratified structure of the base space). }
\end{center}
\end{figure} 
Of course, any horizontal lift still leaves the freedom to have a global transformation, identical at the endpoints of each curve. For any given choice on the particular initial metric $g_1$ representing the initial conformal geometry $[g_1]$, the horizontal  lift $\gamma_H$ of the given path $[\gamma]$ will determine the form of the final metric as well $g_2=\gamma_H(t_2)$.\footnote{In \cite{Teitelboim_path}, Teitelboim seeks to keep the arbitrariness in coordinate system independent at each end. For this purpose, he introduces a second interval in the path integral, with different gauge conditions. This severely complicates the analysis, as now one must also worry about folding (gluing) properties of the path integral. } Of course, not all the paths need end at the same representative, i.e. in principle, for two different base space paths $[\gamma], [\tilde \gamma]$:
$$ \gamma_H(t_2)=g_2\neq \tilde g_2=\tilde\gamma_H(t_2)$$
In fact, this non-trivial holonomy is a representation of the curvature of the connection used to horizontally lift the paths. While it is known that the purely conformal lift has trivial holonomy, the diffeomorphism part usually has a non-trivial one \cite{Gil_Medrano, Freed, Michor_general, Gauge_riem}.

We are interested in a wave-function of the Universe, a la Hartle-Hawking but here with an initial state given by the orbit $[g_o]$, of the least dimensional stratum of $\mathcal{Q}/\mathcal{C}$. Choosing any metric $g$,  schematically we have:
\be \label{wave-function_scheme} \Psi(g):=\int \mathcal{D}f \int_{f^*g_o}^{g} \mathcal{D}\gamma \exp{i S[\gamma_H(f^*g_o, g)]/\hbar}
\ee
  where here $f\in{\mbox{Diff}}$ acts on an arbitrary representative of $g_{ab}^o$ and is integrated over with some measure (which we discuss).  A Haar measure is not required here;  unlike the standard case, we are not doing a group averaging procedure: each path on the base space corresponds to at most one $f$. That is, starting from $g$, each horizontal lift of the curve $[\gamma]$ will hit $f^*g^o$ for a unique $f$.  For instance, if the curvature of the field-space connection form is zero, there is no relative holonomy on the fiber for two paths  $\gamma_1, \gamma_2$, interpolating between $g$ and (the orbit of) $[g_o]$. Thus all the lifts for the paths would end up in a single height, a single representative, $f^*g_o$. Then the path integral over $\mathcal{D}f$  will acquire a functional delta, cancelling the  integral. This is what happens with the purely conformal transformations, since they are Abelian and our choice above \eqref{conf_gf} has no associated curvature \cite{Gil_Medrano}. Interestingly, these sort of gauge-fixings avoid a Gribov ambiguity, but only at the cost of dealing with a further functional integral and the effects of field-space curvature. 

The wavefunction, given by $\Psi(g)$,  obeys the respective conservation laws:\footnote{We note that in the case of odd-dimensional Weyl symmetries, there is  no conformal anomaly, and thus the path measure can be suitably made Weyl-invariant in conjunction with the action \cite{Bertlmann}. This is also true in the Hamiltonian setting if the anomalies have a local representation \cite{SD_Weyl}.} 
\be\label{wavefunction_conserv} g_{ab}\diby{\Psi(g)}{g_{ab}}=0, \qquad \nabla_a\diby{\Psi(g)}{g_{ab}}=0\ee
as I will now show. 

\subsection{BRST invariance}\label{sec:BRST}
\paragraph*{The ghost action}

The ghost action is obtained by replacing the infinitesimal parameters in  the gauge variation of the gauge-fixings \eqref{shift_gf} and \eqref{conf_gf}, given by \eqref{equ:all_transf}, by anti-commuting parameters, the ghosts, $\eta^a, \eta$, and contracting them with the anti-ghosts, $\bar\eta_a, \bar \eta$. 
That is, taking the BRST variations (which we define in the next paragraph)
\begin{eqnarray}
\delta_B \pmb{\xi}&=&\dot {\pmb{\eta}}-[\pmb{\xi} ,\pmb{\eta}]\label{xi_BRST}\\
\delta_B \rho &=&\dot \eta-\xi^ a\partial_a\eta -\eta^ a\partial_a\rho \label{rho_BRST}
\end{eqnarray}
where $\pmb{\xi}=\xi^a\frac{\partial}{\partial x^a}$ and the square brackets stand for the usual vector field commutator.  One thus obtains:
 \be\label{pre_ghost}
S_{\mbox{\tiny ghost}}= -i\int dt \int d^3x \sqrt{g}\, \left(\bar\eta_a(\dot \eta^a-[\pmb{\xi} ,\pmb{\eta}]^a)+\bar \eta (\dot \eta-\xi^ a\partial_a\eta -\eta^ a\partial_a\rho)\right)
\ee
 which on the gauge-fixing surface becomes simply:
\be\label{ghost}
S_{\mbox{\tiny ghost}}\hat= -i\int dt \int d^3x \sqrt{g}\, \left(\bar\eta_a \dot \eta^a+\bar \eta \dot\eta\right)
\ee
where the hat indicates restriction to the gauge-fixing surface. The reduction in complexity, when compared to the similar term in the unitary gauge-fixing of ADM gravity, \cite{Teitelboim_path} is considerable.

\paragraph*{BRST transformations}

I mentioned above the BRST variations. They are defined as above for the fields $\xi^ a$ and $\rho$. For the metric,  they are defined in the same way, replacing $v^a$ and $\theta$ by $\eta^a$ and $\eta$. This guarantees that the classical part of the action, i.e. \eqref{classical_action}, remains BRST invariant. For $\eta$ and $\eta^a$, the BRST transformations are defined as follows:
\be\label{BRST_ghosts}
\delta_B \eta^a=\frac12[\pmb{\eta} ,\pmb{\eta}]^a,\qquad  \delta_B \eta=\eta^a\partial_c\eta
\ee
The first equation does not vanish due to the anti-commuting nature of the ghosts, indeed $[\pmb{\eta},\pmb{\eta}]^a=2\eta^b\partial_b\eta^a$. Defining $\zeta^a$  as the time-derivative of $\eta^a$, from the commutation property of the BRST differential with the time derivative:
 \be\label{equ:time_BRST}\delta_B\zeta^a:= \delta_B \frac{d\eta^a}{dt}= \frac{d}{dt}\delta_B \eta^a=\frac{1}{2}([\pmb{\eta},\pmb{\zeta}]^a+[\pmb{\zeta},\pmb{\eta}]^a)=[\pmb{\zeta},\pmb{\eta}]^a~~\mbox{with}~~ \zeta^a=\frac{d\eta^ a}{dt}\ee

    Now, $\delta_B^2$ can be see to vanish when acting on  $\eta^a$ due to the Jacobi identity.  On $\eta$ we have 
$$\delta^2_B\eta=\frac{1}{2}[\pmb{\eta},\pmb{\eta}]^a\partial_a\eta+\eta^a\partial_a(\eta^b\partial_b\eta)$$
which vanishes since  $\eta^a\eta^b\partial_a\partial_b\eta=0$ due to anti-symmetry. 

For the metric we have:  
$$ \delta^2_B g_{ab}=\frac{1}{2}\mathcal{L}_{[{\pmb{\eta}},{\pmb{\eta}}]} g_{ab}+{{\eta}}^a\partial_a\eta\, g_{ab}-\mathcal{L}_{\pmb{\eta}}( \mathcal{L}_{\pmb{\eta}} g_{ab}+\eta\, g_{ab})-\eta( \mathcal{L}_{\pmb{\eta}} g_{ab}+\eta\, g_{ab})
$$ this vanishes since: i)\, $\eta\eta=0$\,,\, ii), $\mathcal{L}_{\pmb{\eta}}(\eta\, g_{ab})={{\eta}}^a\partial_a\eta g_{ab}-\eta\mathcal{L}_{\pmb{\eta}} g_{ab}$ and\, iii) $\mathcal{L}_{\pmb{\eta}} \mathcal{L}_{\pmb{\eta}} g_{ab}=\frac{1}{2}\mathcal{L}_{[{\pmb{\eta}},{\pmb{\eta}}]} g_{ab}$.
The actions on $\dot g_{ab}$ and on the other time derivatives of the fields necessarily vanish from \eqref{equ:time_BRST} and the constituent transformations.  

To see it directly, we require the Jacobi identity for Lie superalgebras. We have that, for ghost number one fields, $x,y$:
\be\label{equ:Jacobi_identity}
[x,[y,z]]=[[x,y],z]-[y,[x,z]]
\ee 
Thus, for example, for $\delta_B^2\zeta^a$:
\be\label{equ:nilpotent_zeta}
\delta_B(\delta_B\pmb{\zeta})=\delta_B([\pmb{\eta},\pmb{\zeta}])=\frac{1}{2}[[\pmb{\eta},\pmb{\eta}],\pmb{\zeta}]-[\pmb{\eta},[\pmb{\eta},\pmb{\zeta}]]=0
\ee
since from \eqref{equ:Jacobi_identity}, for $x=y=\eta^a$, and $z=\zeta^a$:\footnote{Note that the manner in which I wrote the super Jacobi identity, \eqref{equ:Jacobi_identity} doesn\rq{}t depend on the parity of $z$. The full form for general grading is: $[x,[y,z]]=[[x,y],z]-(-1)^{|x||y|}[y,[x,z]]$. } 
$$2[\pmb{\eta},[\pmb{\eta},\pmb{\zeta}]]^a=[[\pmb{\eta},\pmb{\eta}],\pmb{\zeta}]^a
$$

Lastly, we apply the BRST transformation to the transformation of the Lagrange multiplier \eqref{xi_BRST} (equation \eqref{rho_BRST} follows a similar pattern). We have: 
$$\delta_B(\dot{\pmb{\eta}}-[\pmb{\xi} ,\pmb{\eta}])=[ \dot{\pmb{\eta}},{\pmb{\eta}}]-[\dot {\pmb{\eta}}+[\pmb{\xi} ,\pmb{\eta}],{\pmb{\eta}}]-\frac12[[\pmb{\xi} ,[\pmb{\eta},{\pmb{\eta}}]=0$$

 If the constraints are first class and the structure functions are constants,  i.e.  the algebra of constraints is not ``soft", then the BRST charge is of \emph{rank one} and comes in the following form:
\be   Q= \eta^a\chi _a-\frac{1}{2}\eta^b\eta^aU_{ab}^c P_c
\ee
for $\xi$ the first class constraints, $U$ the structure functions,  and $P$ the ghost momenta. The \emph{rank} of a system can be identified with the order of ghost momenta required for constructing a nilpotent BRST charge.
The nilpotent, rank 1 Hamiltonian generator of the corresponding BRST symmetry in the conformal diffeomorphism case is then:
\be\label{BRST_Q}
Q=\int d^3x\left((\eta^a \, g_{ac} \nabla_b \pi^{bc}-P_b\eta^a\partial_a\eta^b )+(\eta \pi -\frac12P\eta^a\partial_a\eta \right)
\ee
with the ghost momenta given according to \eqref{ghost}, by $P_a=\sqrt{g}\bar\eta_a,\, P=\sqrt{g}\bar \eta$. 

 \subsection{Invariance of the path integral.}\label{sec:inv}
   
Here, we will sketch the construction of \cite{Hartle_Halliwell}, showing  that in our case the  constructed wave-function indeed satisfies the conservation equations \eqref{wavefunction_conserv}. The linearity of the constraints greatly simplifies aspects of the proof, however. 

For coordinate variables $q_i$, momentum coordinates $p^ i$ and Lagrange multipliers $\lambda_\alpha$, encompassed by the variables $z^A$, for an action functional $S[q^i,  p^i,\lambda_\alpha]=S[z^A]$ with certain invariances,   Hartle and Halliwell want to show that wave-function constructed by a sum over paths satisfy corresponding constraints.

The sum over paths is
\begin{equation}\label{fix_path}
\Psi(q^{i^{\prime\prime}}) = \int_{\mathfrak{C}} \mathcal{D} z^A \delta (q^{i}(t^{\prime\prime}) - q^{i^{\prime\prime}}) \Delta_C[z^A] \delta[C^\alpha(z^A)] \exp(i S[z^A])
\end{equation}
where $C^\alpha$ are the gauge-fixing conditions, $\Delta_C[z^A]$ are weight factors associated to the gauge-fixing conditions.  $\mathfrak{C}$ is the class of paths being summed over, including the integration over the final values, $q^{i}(t^{\prime\prime})$. It is the surface delta function that ensures that paths end up at the argument of the wave-function, $q^{i^{\prime\prime}}$.\footnote{Note, however, that if one were to insert a second delta function, determining the anchor of the wave-function, the gauge-fixing \eqref{shift_gf} would not be legitimate (although \eqref{conf_gf} would be). One would have to adopt two different sets of gauge-fixings, as is done for GR in \cite{Teitelboim_path} and there further greatly complicate the analysis. This is the reason that one integrates over the orbit of the initial point. } 

Their claim relies on the following assumptions:
\begin{enumerate}
\item Under a given transformation $\delta_\epsilon q^i= \epsilon^\alpha h_\alpha(p_i, q^i)$, for some function $h_\alpha$ depending solely on the $q^ i, p_i$, parametrized linearly by $\epsilon^ \alpha$, the action functional $S[z^A]$ changes at the most by a surface term, i.e. $\delta S=[\epsilon^\alpha F_\alpha(p_i, q^i)]^{t^{\prime\prime}}_{t^\prime}$. 
\item The class of paths $\mathfrak{C}$ is invariant under the given transformation. 
\item The path integral is independent of the gauge-fixing conditions, at least for a class 
$$C_\epsilon^\alpha[z^A]=C^\alpha[z^A+\delta_\epsilon z^A]$$
\item The combination of measure $\mathcal{D} z^A$ and gauge-fixing weight factor transform according to 
$$ \mathcal{D} z^A\Delta_C[z^A]\rightarrow \mathcal{D}z^A \Delta_{C_\epsilon}[z^A]
$$
\item Path integrals weighted by functions of $q^i, p^i$ on the final surface are equal to (appropriately ordered) operators acting on the wave-function $\Psi(q^{i^{\prime\prime}})$.  In other words, for a given $H(p_i, q^i)$, 
$$ \int_{\mathfrak{C}} \mathcal{D} z^A H(p_{i}(t^{\prime\prime}, q^{i}(t^{\prime\prime}))\delta (q^{i}(t^{\prime\prime}) - q^{i^{\prime\prime}}) \Delta_C[z^A] \delta[C^\alpha(z^A)] \exp(i S[z^A])=H(i\diby{}{q^{i^{\prime\prime}}}, q^{i^{\prime\prime}})\Psi(q^{i^{\prime\prime}})$$
\end{enumerate}
Consequences or not of the path integral construction, these are taken to be the minimal criteria under which invariance of the wave-function follows.

 For a large set of theories (including GR), under the action of the symmetries related to the constraint $Q_\alpha(q,p)$, the associated boundary term for the variation becomes 
$$\delta S=\left[\int d^3 x \epsilon^\alpha(p_i\diby{Q_\alpha}{p_i}-Q_\alpha)\right]^{t_f}_{t_i}$$
which thus vanishes for linear generators, but does not for the scalar Hamiltonian generator of GR.  A non-trivial presence of the boundary term of item 1 slightly complicates the application of the Fradkin-Vilkowisky theorem  to the path integral, as it becomes dependent on a manipulation of  non-BRST invariant boundary conditions (see sec III of \cite{Hartle_Halliwell}).  The further issue which complicates the verification of invariance in ADM is that its commutation relations do not form a Lie algebra, and therefore one cannot interpret the weights of item 4 as the simple Fadeev-Popov determinants, which are the Jacobians which emerge for integration along the gauge-fixing surfaces, and thus arises the requirement 4. There, one must use the full apparatus of the BFV approach. In our case,  all these complications are avoided, items 3 and 4 become the same, and we will not require item 5 at all. 

Let us reproduce our version of the items. 
\begin{enumerate}
\item[1\rq{}] As given in \eqref{equ:all_transf}, $\delta{g}_{ab}(x)=\mathcal{L}_{\pmb{\epsilon}}{g}_{ab}(x)+\epsilon g_{ab}(x)$ the action functional $S[z^A]$ does not change at all, i.e. $\delta S=0$.  This simplifies many aspects of the proof.

\item[2\rq{}] The class of paths $\mathfrak{C}$ is invariant under the given transformation. Here the class of paths are all those in $\mathcal{Q}$. Since the group acts intrinsically in configuration space (which is how we found it), this class is invariant.\footnote{In the case of Hartle-Hawking, the original definition \cite{Hartle_Hawking} only uses paths in $\mathcal{Q}$ which have $\det{(g)}>0$. This is, of course, not a spacetime condition, and thus slice-dependent. So it is not clear to me why such a class of paths is invariant under the assumed symmetry. Of course, this presents no problem in the usual minisuperspace approximation.\label{footnote} }
\item[3\rq{}] The path integral is independent of the gauge-fixing conditions.
\end{enumerate}
where we have joined the previous items 3 and 4, and there is no substitute for item 5. 

The only item above which still requires explanation is item 3. The easiest way to see it is first to notice that the manner in which we have arrived at the ghost action \eqref{ghost} is predicated on its interpretation as a functional determinant (I sketch the proof of this part in appendix \ref{app:FP}). In the presence of a principal fiber bundle structure, this is easy to prove (see \cite{Jaskolski} for a nice geometrical interpretation and proof using the PFB structure).

    Using the Nakanishi-Laudrup trick, we can join the gauge-fixing action \eqref{S_gf} with the ghost action \eqref{pre_ghost}, in the following way (I\rq{}ll do it for the diffeomorphisms, the conformal transformations follow suit): let $B_a$ be a Bosonic (i.e. commuting, or of ghost number zero) variable such that $\delta_B \bar \eta_a= B_a$ (and we define $\delta_B B_a=0$). Then, the term:
\be
\Theta=\bar \eta^a(\frac12 \sigma B_a-\xi_a)
\ee
has a BRST variation of the form: 
\be\label{equ:gf_ghost}
\delta_B\Theta=B_b g^{ab}(\frac12 \sigma B_a-\xi_a)-\bar\eta^a(\dot \eta_a-[\pmb{\xi}, \pmb{\eta}]_a)
\ee
which is what we want to add to the classical action. But now it is easy to see that if we solve the equations of motion for the auxiliary variable $B_a$, we obtain $\sigma B_a=\xi_a$. Inputting this back into \eqref{equ:gf_ghost} we obtain two terms: \eqref{pre_ghost} and \eqref{S_gf}. From the nilpotency of $\delta_B$, the entire gauge-fixed action, 
$$S_{\mbox{\tiny cl}}+S_{\mbox{\tiny gf}}+S_{\mbox{\tiny gh}}=S_{\mbox{\tiny cl}}+\delta_B\Omega$$ is BRST invariant. 

For the measure, as mentioned, we are secretly using the Liouville measure and projecting to configuration space. But, under a BRST transformation, the Liouville measure transforms by a total derivative term which is just a canonical transformation (see e.g. appendix A of \cite{Hartle_Halliwell}). 

Finally, given these items, we proceed to prove the invariance of the wave-function \eqref{fix_path}. 
The only term that transforms in \eqref{fix_path} under a gauge transformation is the delta function, in our language $\delta(g_{ab}(t^{\prime\prime})-g_{ab}^{\prime\prime})=\delta(g_{ab}^1-g_{ab}^2)$, which transforms  under an infinitesimal diffeomorphism acting on $g_{ab}^1$. 
But we have, shifting variables, according to the standard property of functional deltas, \eqref{delta_identity}, up to first order,
$$\delta(g_{ab}^1 +\mathcal{L}_\epsilon g_{ab}^1 - g_{ab}^2)=\delta(g_{ab}^1  - (g_{ab}^2-+\mathcal{L}_\epsilon g_{ab}^2))\left(\det{(\mbox{Id}+\diby{}{g_{ab}^1(y)}\mathcal{L}_\epsilon g_{ab}^1(x))}\right)^{-1}
$$
The identity for $\delta$ implicitly uses a redefinition of the integration variable to produce the determinant. Now $\det(1 + \epsilon) = 1 + tr(\epsilon)$, where the trace is an integration over $x=y$ and summation over internal indices. Since the structure constants of our group are traceless,\footnote{This is in essence the same reason the Fadeev-Popov procedure works simply in our case. See appendix \ref{app:FP}. This is not the case with the standard ADM algebra.} we get
\begin{align} 
\delta(g_{ab}^1 +\mathcal{L}_\epsilon g_{ab}^1 - g_{ab}^2) =  \delta(g_{ab}^1  - g_{ab}^2)- \int \rd^3 x \mathcal{L}_\epsilon g_{ab}^2(x) \frac{\delta}{\delta g_{ab}^2(x)}\delta(g_{ab}^1  - g_{ab}^2)\label{bla}
\end{align}

Finally, we will implement this result in \eqref{fix_path}, by shifting the integration variables, and using 1\rq{}, 2\rq{} and 3\rq{}, above. Noticing that none of the variables depend on (the equivalent of) $g_{ab}^2$, and thus we can apply the functional derivative on the rhs of \eqref{bla} to everything:
\begin{align}
 0 ={}& \int_{\mathfrak{C}} \mathcal D g^1_{ab} \int_{M} \rd^3 x  \mathcal{L}_\epsilon g_{ab}^2(x) \frac{\delta}{\delta  g_{ab}^2 (x)} \Big(\delta (g_{ab}^1 - g_{ab}^2) \Delta_C[g^1_{ab}] \delta[C[g^1_{ab}]] \exp(-\sigma S[g^1_{ab}]) \Big)\\
 0={}&- \int_{M} \rd^3 x \epsilon^a\nabla_2^b \frac{\delta}{\delta g_{ab}^2(x)} \Psi[\mathbf{g}_2]
\end{align}
where $\nabla^a_2$ refers to the metric $g^2_{ab}$. 
This shows the second equation of \eqref{wavefunction_conserv} holds. The first follows from the same procedure and is much simpler as  the conformal group is Abelian.\footnote{See \cite{SD_Weyl}, for a study of the appearance of conformal anomalies for the conformal symmetry in the Hamiltonian setting appropriate to this foliation-preserving group, $\mathcal{C}$. }

\paragraph{Perturbation theory}

In fact, for theories that have preferred foliations, a similar gauge-fixing of the path integral is performed in \cite{Path_foliations}. There, one chooses gauge-fixings of the space-time diffeomorphisms only through gauge-fixings of the perturbations of the lapse and shift, not for the metric and its perturbations. That is also what is being done here: conditions on the propagation of coordinates are being chosen, rather than the coordinates themselves. The gauge-fixing terms become much simplified --- one can have just terms that set the perturbations of the shift to zero for example --- as opposed to choosing e.g.  harmonic coordinate conditions for the full perturbations of the metric tensor. 

To actually perform calculations for perturbation theory, we need to find a background of all the fields (including Lagrange multipliers if there are any), and then perform the second variation of the gauge-fixed action. The problem that shows up here, and not elsewhere,  is that for perturbative renormalizability we need to choose a background more amenable to the constraints of the theory than the usual Minkowski background; a  Minkowski background is not directly suitable for the calculations, since it is conformally flat and thus has vanishing Cotton tensor.

As mentioned before, the physical degrees of freedom of the conformal geodesic theory are the same as in general relativity: both have the transverse traceless gravitational momenta, as per equation \eqref{equ:constraints}.  But we need to find a linearized version of the theory if we would like to obtain a linearized wave equation, as in general relativity. Ideally, we would have  an approximation to the calculation on the full Bianchi IX background.  This will be pursued in a further paper. 

For now, to get an idea for the structure of the equations, we can use the preferred configuration $g_o$, the 3-sphere metric,  to construct a density which is not affected by conformal transformations. This allows us to write a different action: 
$$S= \int dt \int_M
 d^3 x\,d^3x (\dot g_{\mbox{\tiny H}}^{cd}\dot g^{\mbox{\tiny H}}_{cd}+\Lambda) (\sqrt{g}\, \Xi\,+a\sqrt{g_o})=\int dt\int d^3x\, TV
 $$
 where I wrote the action as the multiplication of a kinetic term $T$ and potential one $V$. Now it is possible to calculate the Euler-Lagrange equations of motion around a spherically symmetric background, i.e. $g_{ab}\rightarrow g^o_{ab}+\epsilon h_{ab}$, with $g^o_{ab}$ having radius of curvature $r$. It is easy to see that $\frac{\delta{V}}{\delta g_{ab}}_{|g_{ab}=g^o_{ab}}=0$. 
 After some algebra, one can evaluate the only surviving terms from the Euler-Lagrange equation:
 \begin{eqnarray*} \ddot h_{ab}(x)&=\left.\frac{g}{a g_o} \int d^3z\int d^3 y g^{kc}g^{id}\frac{\delta C_{ki}(z)}{\delta g_{ef}(y)} h_{ef}(y)\frac{\delta C_{cd}(z)}{\delta g_{ab}(x)}\right|_{g_{ab}=g^o_{ab}}\\
 ~ &=\frac{1}{ar^4}\left( 4\nabla^2 h_{ab}-\frac{7}{4}r^2\nabla^4 h_{ab}+r^4\nabla^6 h_{ab}\right)
 \end{eqnarray*}
 This wave equation has the interesting property that it possesses higher order spatial derivatives. This will be a general property of specific solutions to the equation \eqref{equ:eom_pi}. It provides modified dispersion relations that may be fruitful in regularizing gravitational divergencies, as is the case of Horava-Lifschitz \cite{Horava}. 

The study of perturbation theory around a Bianchi IX static solution has only just begun. For future work, we can follow the algorithm set in \cite{Barvinsky_hor} as applied to this theory.

\section{Conclusions}
\subsection{Summary}

\paragraph*{Classical}
I started the paper looking for gravitational theories which admit a reduced configuration space in metric variables. That is,   superspace $\mathcal{Q}/\mbox{Diff}(M)$ is not the reduced space of physical configurations for GR; one still has an action of refoliations. Thus the question arises: what is the maximal local symmetry group acting on metric configuration space that admits a respective \lq\lq{}superspace\rq\rq{}? Such a group represents symmetries compatible with a quantum mechanical transition amplitude between physical observables, for which no \lq\lq{}Problem of Time\rq\rq{} emerges. 

The answer to the question posed above is: conformal diffeomorpshims. Under the action of these transformations, the reduced quotient configuration space formed  is called \lq{}conformal superspace\rq{}. Standard dynamical approaches to GR often bear hints of this symmetry: it is apparent in numerical approaches (it is used to interpret LIGO data \cite{Pretorius_binary, BSSN_binary}); it is used for studying the initial value problem of GR \cite{York, 3+1_book}, and it is used to separate out the physical degrees of freedom of GR from the gauge ones \cite{IsMa1982}. 

    Shape dynamics (\cite{SD_first}, see \cite{Flavio_tutorial} for a review) is a theory that embodies these symmetry principles. However, it is engineered in such a way as to closely match GR dynamics, thereby acquiring a non-local Hamiltonian. Therefore, starting just from the same symmetry principles and assumptions of locality, it becomes of paramount importance for the entire spatially relational approach (to which shape dynamics belongs) to investigate the existence and properties of this theory space. This investigation should be done even if merely to rule out models other than shape dynamics belonging to this class, in which case, more confidence should be invested in the only existing completely  first principles approach to  the construction of shape dynamics (given in \cite{SD_construction}):  \lq{}symmetry-doubling\rq{} (which could also inform extensions of gauge groups for particle physics). 

  For these reasons, we then went on to investigate the almost local (possessing a square root of derivative terms) gravitational actions that were compatible with local spatially conformal theories. 
This field and symmetry content guarantee that the emerging theory has the correct transverse traceless gravitational physical degrees of freedom,  a desirable result, vis a vis the great struggle to modify gravity while keeping its two degrees of freedom intact.   Lastly, one can use the structure of conformal superspace to establish a unique preferred orbit. This is, in a very precise sense, the most homogeneous of all configurations, and it has a preferred geometrical status on reduced configuration space. This preferred point serves to establish  a preferred initial point in the path integral kernel, $\phi^*=g_o$, as above, thereby defining the static wavefunction in a similar fashion as  the Hartle-Hawking boundary conditions \cite{Hartle_Hawking}.

 This principled derivation of the theory sets it apart from the standard modifications of gravity already at a classical level; it is not ad hoc  (see e.g. \cite{Horndeski})  and it contains no spin-0 (or scalar) gravitational degree of freedom \cite{Horava}. Regarding the quantum mechanical theory, departing from the same principles, one would not be able to define \textit{different}, natural, \lq{}initial conditions\rq{} for the wave-function of the Universe, $\Psi(g)$.  Because the nested boundary structure in conformal superspace has a unique, asymmetric, topological structure which strongly suggests the present boundary conditions.  For standard GR there is more ambiguity (see e.g. \cite{Hartle_Hawking, Linde, Vilenkin}) because one does not have access to the true configuration degrees of freedom (see also footnote \ref{footnote}).

Further addressing the scientific merit of the theory, it seems very vulnerable to falsification, which is a positive characteristic. Minimal coupling to vector fields yields a $U(1)$ gauge theory with hyperbolic equations of motion, a surprising consequence from our starting point. For cosmology, one could study what is known for Bianchi IX  (see e.g. \cite{Wiltshire_intro}) and  the modification affected by breaking the adiabatic limit (see last paragraphs of section \ref{sec:matter}).   It is  encouraging that the Yang-Mills form of the Hamiltonian for (Lie-algebra valued) one-forms emerges very naturally, from the requirements of conformal covariance and conformal weight matching, in \eqref{equ:vector_Hamiltonian}. The emergent hyperbolic equations for YM mean that the electromagnetic sector would have a  light cone. Nonetheless, as can be seen from equation \eqref{full_EM}, there are discrepancies at lower derivatives, of the form 
$$ -\frac13  \frac{\alpha_{\mbox{\tiny v}}\beta_{\mbox{\tiny v}}}{\Xi ^{2/3}}\nabla_{[a}A_{b]}\partial^b \ln \Xi -\dot A_a\left(\frac{\dot g}{2}+\frac{1}{3 }\frac{d}{dt} \ln \Xi\right)
$$ This shows some hope in modifying quantum cosmology in a regime-dependent manner, where inhomogeneities in space and time affect the propagation of light. The recent results on GW170817 \cite{GW_NS}, show that the light-speed and the gravity-speed can differ at most by 1 part in $10^{15}$. It would be extremely interesting  to study when inhomogeneity terms become relevant in standard Bianchi IX type of cosmologies for the present construction, and if bounds on homogeneity are compatible the bounds on the discrepancy of speeds.  I leave this for further study. 

  I also leave for further study an analogue of the Birkhoff theorem. Nonetheless, it is possible to recover, through a simple coupling of a scalar field in approximate spherical symmetry (again regulated by the magnitude of $\partial^b \ln \Xi$)  and static metric background, a Schwarzschild metric in isotropic coordinates (thus with no interior of the black hole). This is analogous to how one discovers these solutions (and proves a Birkhoff theorem) for shape dynamics \cite{SD_birkhoff} (see also \cite{Flavio_single} for a criticism of the boundary conditions used there).  It is important to note that departures from stationary GR, in both the propagation of light and in the Schwarzschild solution,  is run by the same parameter, $\partial^b \ln \Xi$.
  
 % The distinct  extra coupling of all matter fields and gravitons to the geometry through multiplication by powers of $C^{ab}C_{ab}$ can give an extra, dynamical  hierarchy to the different interactions.  It seems that all of our couplings to the gravitational sector have to be in some sense non-minimal, meaning that different regimes of gravity can suppress or amplify different types of interactions.\footnote{ In effect, such non-minimal coupling terms seem to also emerge in the usual renormalization group flow of the Einstein-Hilbert action when coupled to the standard model sector \cite{Astrid_matter}.}. 

\paragraph*{Quantum}
With regards to technical questions surrounding quantum gravity, the theory seems to have several advantages. To be specific, they  are seven-fold: {\bf i)} it is second order in time derivatives (in space-time covariant field space, Lagrangians extending general relativity require more time derivatives, creating possible problems for unitarity, since the propagators around a flat background acquire imaginary poles), {\bf ii)} the theory explicitly maintains two, transverse traceless, propagating degrees of freedom per space point (this is an issue with other theories, such as Horava-Lifshitz and Horndeski-type theories), {\bf iii)} The treatment of the path integral is very simplified because of simple constraint algebras (for  GR, difficulties introduced by the refoliations constraint make this treatment much lengthier; it encompasses parts of both \cite{Teitelboim_path, Hartle_Halliwell}).  The theory possesses  simple BRST symmetry (hard to achieve for the ADM Dirac algebra for example, which requires a BFV treatment \cite{Hartle_Halliwell}, see appendix \ref{app:FP} for reasons why). {\bf iv)} There is no Gribov problem for the horizontal lift gauge-fixing (a non-perturbative problem for basically all non-abelian theories based on space-time fields \cite{Singer_Gribov}), {\bf v)} superposition can readily be made sense of, simply as the interference between (coarse-grained) paths in configuration space (see \cite{QG_deco}), there is no need to define superposition of causal structures,  %{\bf vii)} for general theories of this sort, conservation of probability ensues (again, see \cite{path_deco}),  
 {\bf vi)}  the linearized equations of motion  contain higher order spatial (but not temporal) derivatives. This property is extremely desirable in the context of regularizing ultraviolet divergencies (as is done in e.g. \cite{Horava}), a topic we will explore in further work.     The study of perturbation theory around a Bianchi IX static solution has only just begun. For future work, we can follow the algorithm set in \cite{Barvinsky_hor} as applied to this theory. 
 
   Indeed, I believe that these features already make these theories  extremely interesting as a toy model, irrespective of having the correct classical limit,\footnote{Most quantum gravity theories have \emph{also} the problem of matching known physics. String theory doesn\rq{}t  \textit{explicitly} recover the full Standard Model and Loop Quantum gravity and spin foams don\rq{}t recover realistic Einstein space-times in the appropriate limits. Asymptotic safety is on better grounds on this respect, since it is more in line with an effective field theory  approach.} and many of the features studied here are clearly applicable to shape dynamics.

\section*{ACKOWLEDGMENTS}
 This research was supported  by Perimeter Institute for Theoretical Physics. Research at Perimeter Institute is supported by the Government of Canada through Industry Canada and by the Province of Ontario through the Ministry of Research and Innovation.

\appendix
\section*{APPENDIX}

\section{Gauge theory in Riem}\label{app:horizontal}
The results of this appendix are used  in section \ref{sec:past}.

 \subsection{Slice theorem for Riem}\label{app:slice}
 \begin{defi}[Slice]\label{def:slice}
Let $G\supset I_x=\{g\in G~|~g x=x\}$ be the isotropy subgroup of $G$ at $x$.  A slice at $x\in \mathcal{X}$ for the action of $G$ is a submanifold $S_x$ such that  $x\in S_x\subset\mathcal{X}$ and:
 \begin{itemize}
 \item if $g\in I_x$, then $gS_x=S_x$;
 \item if $g\in G$ and $(gS_x)\cap S_x\neq \emptyset$, then $g\in I_x$; 
 \item there is a map $\mu:G/I_x\rightarrow G$, called a local cross-section, defined in a neighborhood $U$ of the identity of $G$, such that the map $\mathcal{F}:U\times S_x\rightarrow \mathcal{X}$, defined by $\mathcal{F}(g, y)=\mu(g)y$ is a diffeomorphism onto a neighborhood $\mathcal{U}$ of $x$. 
 \end{itemize}
 \end{defi}
\begin{cor}\label{cor}
From the existence of a slice $S_x$ defined as above, it is  easy to show that  for a given neighborhood $\mathcal{U}$ of $x$, \textbf{i)} for $y\in \mathcal{U}\cap S_x$ then 
$I_y\subset I_x$ (follows from the second item in definition \ref{def:slice}); and \textbf{ii)} for $y\in \mathcal{U}$ then $I_y$ is conjugate to $I_x$ (follows from successively applying the third property of definition \ref{def:slice} and item \textbf{i)}).

From a slice, we also trivially obtain the local product structure generically, i.e. for points in $\hat{\mathcal{X}}=\{x\in \mathcal{X}~|~I_x=Id\}$, we have $\mathcal{U}\simeq S_x\times G$, with $\mathcal{U}$ being a proper subset of $\hat{\mathcal{X}}$ containing $x$. It follows that the quotient $\hat{\mathcal{X}}/G$ has a manifold structure.  
\end{cor}

 Ebin and Palais have shown that $\mathcal{Q}$ has a local slice \cite{Ebin, Palais}.  They did this through the use of the normal exponential map to the orbit along a given point. First, one establishes the following: 
  Let the configuration space in question, $\mathcal{Q}$, be the field space of Euclidean d-dimensional metrics:
$$g_{\mu\nu}\in C_+^{\infty}(T^*M\otimes_S T^*M):=\mathcal{Q}$$
Now, for two field-space tangent vectors at $g_{\mu\nu}$ (usually denoted by $\delta_1 g_{\mu\nu}, \delta_2 g_{\mu\nu}$ \cite{Wald_Lee}), 
  $$u, v\in C^{\infty}(T^*M\otimes_S T^*M)\equiv T_g\mathcal{Q},$$
  the orbit under the gauge-symmetries of GR --- the diffeomorphisms \mbox{Diff}(M) --- is given by
$$ \mathcal{O}_g=\{ f^*g_{\mu\nu}| f\in\mbox{Diff}(M)\},~X\in C^\infty(TM), ~~ \mathcal{L}_X g_{\mu\nu}\in T_g\mathcal{O}_g=:V_g$$
     For   a supermetric    $$ \langle u, v\rangle_g:=\int d^4x\sqrt{g}\, u_{\mu\nu}\, g^{\mu\rho}\,g^{\nu\sigma}\,v_{\rho\sigma}$$ we can find the space orthogonal to the orbit,  $V_g^\perp$. Indeed, 
$$h_{\mu\nu}\in V_g^\perp\Rightarrow \nabla^\mu h_{\mu\nu}=0.$$
Thus, from the form of $V_g$ and \eqref{equ:supermetric},  
\be\label{vertical_perp}V_{\bar g}^\perp=\{u^{ab}\in T_{\bar g}\mathcal{Q}~|~\bar\nabla_a u^{ab}=0\}
\ee 

For a generic $v_{\mu\nu}\in  T_g\mathcal{Q}$, 
$$v_{\mu\nu}=v^\perp_{\mu\nu}+\mathcal{L}_X g_{\mu\nu}$$
where $X^a$ given by solution of
\be\label{elliptic}\nabla^\mu(\nabla_{(\mu}X_{\nu)})=\nabla^\mu v_{\mu\nu}\ee
 which is a 2nd order pde, with a unique solution. This then leads to a unique decomposition:
  $${v_{\mu\nu}^\perp}=v_{\mu\nu}-\mathcal{L}_X g_{\mu\nu}$$
 
This is true infinitesimally. But then, for an arbitrary given metric $\bar{g}_{ab}$, the orbits $\mathcal{O}_{\bar g}$ were shown to be embedded submanifolds, which therefore have a well-defined tubular neighborhood.  Given the tangent of the orbit, $T_{\bar g}\mathcal{O}_{\bar g}=:V_{\bar g}$ and a $G$-invariant supermetric (e.g. \eqref{equ:supermetric} with $\lambda=0$), one can then define the normal exponential map: $\mbox{Exp}_{\bar{g}}:W\subset V_{\bar g}^\perp\rightarrow \mathcal{Q}$, which was shown to be a local diffeomorphism onto its image for a given open set $W$. 
 By then showing that the tubular bundle around this orbit was locally diffeomorphic to $\mathcal{Q}$, one has, for $g_{ab}\in \pi^{-1}(\pi(\mbox{Im}(\mbox{Exp}_{\bar{g}}(W)))$, a unique $f_g\in \mbox{Diff}(M)$ such that 
$$ f^*_g g_{ab} =\mbox{Exp}_{\bar g}(w_g)
$$ for a unique $w_g\in W$, 
$$w_g=\mbox{Exp}_{\bar{g}}^{-1}(f_g^*g_{ab})$$  Thus 
 $$g_{ab}=f^*_g(\mbox{Exp}_{\bar{g}}(w_g))$$
Furthermore, for $g^2_{ab}=h^*g^1_{ab}$, we then have $f_{g^2}=h^{-1}\circ f_{g^1}$. Thus $w_g=w_{[g]}$ and for any $\tilde{g}_{ab}\in [g_{ab}]\subset\pi^{-1}\pi(\mathcal{U})$, the section is given by 
\be \chi(\pi(\tilde g_{ab}))= \mbox{Exp}_{\bar{g}}(w_{[\tilde{g}]})
\ee In more heuristic terms, $\chi$ takes any  metric along the orbits and translates it along the orbit until it hits the orthogonal  exponential section at the height of $\bar{g}_{ab}$. This intersection gives us the value of $\chi$ for the given equivalence class.

\subsection{Principal fiber bundles and connections}\label{app:PFB}
{A principal fiber bundle is a manifold $P$, on which a Lie group $G$ acts freely: $P\times G\rightarrow P$,  here we will assume it acts on the left,  $(p, g)\rightarrow g\cdot p=L_g(p)$. The space $\{g\cdot p\,~|~g\in G\}$ is called the the fiber through $p$. Moreover,  $P$ is assumed to have a locally trivializing section, but we will not need these details here. }

Given a slice theorem for the generic subspaces where $I_x=Id$, we can form such a principal fiber bundle, and make use of all the structure that comes with it. Given $\mathfrak{v}\in \mathfrak{g}$,  where $\mathfrak{g}:=T_{\mbox{\tiny Id}}G$, we define the fundamental vector field associated to it as  $T_pP\ni\mathfrak{v}_p^\#:=\frac{d}{dt}_{|t=0}\exp{(t\mathfrak{v})}\cdot p$. The vertical subspace  $V_p\subset T_pP$ is the tangent space to the fiber at $p$, i.e.  $V_p=\mbox{span}\{\mathfrak{v}_p^\#\,~|~\mathfrak{v}\in \mathfrak{g}\}$.  A horizontal distribution is a smooth equivariant tangential distribution which complements the vertical spaces, i.e  $H_p\subset T_pP$ such that:
\be\label{equ:hor_conds}  \mathbf{i)}\, ~ H_p\oplus V_p=T_pP \,~\, \mbox{and} \,~\,   \mathbf{ii)}\, ~ L_g^*(H_p)=H_{g\cdot p}\ee
 
   Such horizontal spaces are uniquely identified by a connection form on $TP$ -- a $\mathfrak{g}$-valued one-form $\omega$, satisfying:
   \be\label{equ:hor_conn_conds}  \mathbf{i)}\,~ \omega(\mathfrak{v}_p^\#)=\mathfrak{v} \,~\, \mbox{and} \,~\,   \mathbf{ii)}\, ~ L_g^*\omega=\mbox{ad}(g)\omega\ee  The identification is obtained by the vertical projection, for a given $X_p\in T_pP$, it is the fundamental vector field (or vector tangent to the fiber)  $\hat{V}_p(X_p)= (\omega(X_p))^\#$. The horizontal projection  is its complement, $\hat{H}_p=\mbox{Id}-\hat{V}_p(X_p)$. 

We can rewrite the property corresponding to ii) of \eqref{equ:hor_conn_conds} for a field-dependent transformation as (see \cite{Aldo_HG}): 
     \be\label{varpi_tranfo}\delta_{\xi}(\omega(u))^a=\delta \xi^a -[\omega(u), \xi]^a\ee 
     for $u$ a vector at $T_pP$. 
\paragraph*{Connections in Riem.}
  For this geometric approach to gravitational theories proposed here, one can think of configuration space as an infinite-dimensional principal fiber bundle $P=\mathcal{Q}$, foliated by the group orbits of the diffeomorphisms $G=\mbox{Diff}$, acting by pull-backs (and scalar transformations, acting by pointwise multiplication).\footnote{This is not strictly true, because metrics with isometries possess non-trivial stabilizer subgroups of the diffeomorphisms, one obtains a  stratified manifold for the quotient $\mathcal{Q}/$Diff (see \cite{Fischer}). However one can make sense of this as a principal fiber bundle if one considers only the diffeomorphisms which fix one point of $M$, and a linear frame on it. \cite{Giulini_geometrodynamics}.}  

  In this case, best-matching, or the gauge connection in Riem,  is given by a linear action $\xi:TP\rightarrow\mathfrak{g}$, where $\mathfrak{g}$ is the corresponding (infinite-dimensional) Lie algebra (for instance $C^\infty(TM)$ with the Lie bracket for diffeomorphisms) and  $P=\mathcal{Q}$.

 As in usual formulations of gauge symmetry in principal fiber bundle language, this gives a projection $\hat H: \dot g_{ab}\rightarrow \dot{g}_{ab}^{\mbox \tiny H}$ into a `horizontal' component of the metric velocity (see \cite{Bleecker} for a quick introduction to principal fiber bundles). That is, in the Lagrangian formulation, a ``preferred shift" defining an equilocality relation -- called the `best-matched coordinates -- emerges from a connection form (see \cite{Gauge_riem} for details on how these objects are constructed in this infinite-dimensional context).

In plain words, the role of the connection form (i.e. of best-matching) is to project out the pure `coordinate-change' component of a metric velocity. The inverse of the (conformal) thin-sandwich differential operator (see \cite{Maxwell}) is obtainable by such a form which projects out orthogonal components to the fibers, defining $\hat H$. If a given supermetric is positive-definite and covariant with respect to the gauge transformations,  we can define $\hat H$ by the projection orthogonal to the fibers. 

This is what is done for instance in \cite{Ebin} (see  also \cite{FiMa77}), using the canonical (positive-definite) supermetric in Riem,  $G_0^{abcd}:=\frac{1}{2}(g^{ac}g^{bd}+g^{bc}g^{ad})$ (i.e. the canonical supermetric with zero DeWitt parameter). Since fundamental vector fields at $g_{ab}$ (i.e. tensors tangent to the fibers) are given by $\mathcal{L}_\xi g_{ab}=\nabla_{(a}\xi_{b)}=\nabla_a\xi_b+\nabla_b\xi_a$, one obtains that horizontal subspaces are the ones that obey $\nabla^a\dot{g}_{ab}=0$. The vertical projection is given by the vector field-valued one form (i.e. a functional acting linearly on metric velocities with values on the Lie algebra of the diffeomorphisms) $\xi^a$ such that:
\be\label{equ:connections_diff}
\nabla^a(\dot{g}_{ab}-\mathcal{L}_{\xi[\dot g;x)} g_{ab})=0
\ee
i.e. it is defined by the horizontal projection being orthogonal to the fibers according to the supermetric. Because the supermetric is positive definite, we note that  $H_p\cap  V_p=0$, so that, barring metrics with non-trivial isometry group (for which the principal fiber bundle picture needs mending), 
\be\label{equ:ortho_diff}
\nabla^a(\mathcal{L}_{\xi[\dot g;x)} g_{ab})=(\nabla^2\delta^a_b+R^a_b){\xi[\dot g;x)}_a-\nabla_b(\mbox{div}({\xi[\dot g;x)}))=0
\ee
only for ${\xi[\dot g;x)}_a=0$, which is the property corresponding to $\omega(\mathfrak{v}_p^\#)=\mathfrak{v} $.\footnote{ For transverse vector fields, $\nabla_b(\mbox{div}(\xi))=0$, the operator acting on $\xi^a$ corresponds to the operator used in works in asymptotic safety to redefine the vector fields and get rid of unwanted Jacobians, very similarly to what we have done with our horizontal lift in the main text.}

Furthermore, since the supermetric is equivariant wrt to the action of the symmetry group, we automatically obtain that the horizontal projections obey \eqref{equ:hor_conds} and therefore the connection form obeys \eqref{equ:hor_conn_conds}.  
These properties (which will also hold for the conformal connection defined below) guarantee that this functional shift vector transform as \eqref{equ:Lagrange_transfs}. 

\paragraph*{Conformal diffeomorphism connections given by a supermetric in Riem.}
In \cite{Gauge_riem}, one uses the supermetric:
\be\label{equ:conf_supermetric} (v, w)_g:=\int d^3 x\sqrt{g}\, \Xi\, G_0^{abcd}v_{ab}w_{cd}
\ee
where the vectors $v_{ab}$ are taken to transform as $v_{ab}\rightarrow e^{4\rho}v_{ab}$ under conformal transformations $g_{ab}\rightarrow e^{4\rho}g_{ab}$, for $\rho\in C^\infty(M)$ a scalar function.  The supermetric \eqref{equ:conf_supermetric} is positive definite and is equivariant with respect to conformal diffeomorphisms.
The integrand in \eqref{equ:conf_supermetric} is thus invariant under conformal transformations.
Thus, using \eqref{equ:conf_supermetric}, one forms a genuine gauge connection from the orthogonal projection, 
 \be\label{corrected}\dot g_{\st{H}}^{ab}=(\dot g^{ab}-\nabla^{(b}\omega_g^{a)}[\dot g;x)+\omega_g[\dot
g;x)g^{ab})\ee This equation defines $\dot g^{\st{H}}_{ab}$ as  a generalized `transverse traceless' metric velocity,\footnote{In fact, upon a Legendre transformation, the conditions imposed on the horizontal projection become $\nabla_a\pi^{ab}=0$ and $\pi^a_a=0$, the transverse traceless conditions. } corrected by infinite-dimensional gauge connections on Riem, $\omega[\dot g]^a$ and $\omega[\dot
g]$, along the diffeomorphism and conformal fibers respectively, which project the velocities into the space orthogonal to the conformal diffeomorphism fibers, as we did before just with the diffeomorphisms. We note that here too, the horizontal projections obey \eqref{equ:hor_conds} and therefore the connection form obeys \eqref{equ:hor_conn_conds}, and these properties guarantee that this functional shift vector transform as \eqref{equ:Lagrange_transfs}, without any need for any ad hoc definitions of transformation properties.   

To be more specific, rewriting the full group of diffeomorphisms as a product of the incompressible diffeomorphisms and the pure dilatations. These are generated by divergenceless vector fields and their complements. In that case, we have, - from an orthogonality condition to the fiber, according to \eqref{equ:conf_supermetric}:
 \begin{align} \nabla_a\left(\, \Xi\,\,v^{a}_{~b}\right)&=0\label{equ:xi_horizontal}\\
v_{ab}g^{ab}-3\rho &=0\label{equ:rho_horizontal}
    \end{align}
Given a metric velocity, $\dot{g}_{ab}\in T_gP$, we want to horizontally correct it, as is the job of the usual covariant derivative:
 \begin{align}\label{hor_corrected} \nabla_a\left(\, \Xi\,(\dot{g}^{ab}+\nabla^{(a}\xi^{b)}+\rho g^{ab}\right)&=0\\
\dot {g}_{ab}g^{ab}-\mbox{div}{(\xi)}-3\rho &=0
    \end{align}
Since we want to invert this for $\xi^a$ and $\rho$, we look at equations \eqref{equ:xi_horizontal} and \eqref{equ:rho_horizontal} as equations to be inverted for $\xi^ a$ and $\rho$ in terms of sources given by the metric and its time derivatives. 

The condition for the existence of an element of $T_g\mathcal{Q}$ which is both orthogonal to a fiber and parallel to it (i.e. is both horizontal and vertical) is that a non-zero $\xi^a$ exist such that
\be\label{equ:ver_hor}
 \nabla_a\left(\, \Xi\,(\nabla^{(a}\xi_{\st{T}}^{b)}\right)=0
\ee
where we are taking the divergence-free Killing form. 
But in our case we the supermetric \eqref{equ:conf_supermetric} is positive definite, and thus the only solution is $\nabla^{(a}\xi_{\st{T}}^{b)}=0=\mathcal{L}_{\xi_\st{T}} g_{ab}$.\footnote{Unlike what is the case for the DeWitt supermetric, for which indeed one can find non-zero elements belonging to the intersection $H_g\cap V_g$.} Thus equation \eqref{equ:ver_hor} only has non-trivial solutions in case the metric has conformal Killing vector fields (since we are taking the transverse part of the vector fields, and so on). 

We thus construct the two operators on $T\mathcal{Q}$, i.e. $\omega_g^a:T_g\mathcal{Q}\rightarrow C^\infty(TM)$ and $\omega_g:T_g\mathcal{Q}\rightarrow C^\infty(M)$, such that:
 \begin{align} \nabla_a\left(\, \Xi\,((\dot{g}^{ab}-\omega_g[\dot g;x)g^{ab})-\nabla^{(a}\omega_g^{b)}[\dot g;x)\right)&=0\label{equ:hor_diff}\\
\frac{1}{3}\dot {g}_{cd}g^{cd}-\omega_g[\dot g;x) &=0\label{equ:hor_conf}
    \end{align}
from which we obtain \eqref{corrected}. 

For the more formal mathematical proof that one can indeed obtain a connection form from this orthogonality criterion, one  needs to show that the criterion can be expressed as the kernel of an elliptic operator, and use the Fredholm alternative in the infinite-dimensional Banach space case. This is done to full detail  in \cite{Gauge_riem}.

Thus for the geodesic action: 
\be\label{new action} S= \int dt\sqrt{\int_M
\sqrt{g}d^3x \, \Xi\,(\dot g^{cd}_{\st{H}}\dot g^{\st{H}}_{cd})}
  \ee
and by virtue of the connection forms in \eqref{corrected}, under a conformal diffeomorphism  $(f, \rho)$, the integrand changes covariantly wrt to the diffeomorphisms $f$ and invariantly with respect to the conformal transformations, as: 
$$f^*\left(\sqrt{g}d^3x\, \Xi\,(\dot g^{cd}_{\st{H}}\dot g^{\st{H}}_{cd})\right)(x)=\left(\sqrt{g}d^3(f(x))\, \Xi\,(\dot g^{cd}_{\st{H}}\dot g^{\st{H}}_{cd})\right)(f(x))$$

    \section{Fadeev-Popov and Lie algebras.}\label{app:FP}
    
   Morally, the Fadeev-Popov determinant is a simple functional extension of the following argument: for a function with a single root, $f(x_o)=0$, the Dirac delta function obeys the identity: 
   \be\label{delta_identity}\delta(f(x))=\frac{\delta(x-x_o)}{|\det{f\rq{}(x_o)}|}\ee Since the integral of $\delta(x-x_o)$ over $x$ gives unity, then
$$|\det{f\rq{}(x_o)}|\int dx \, \delta(f(x))=1$$
The functional setting works morally in the same way. 
% with \be\label{dirac_FP}
%|\det{J\rq{}(\varphi_o)}|\int \mathcal{D}\varphi\, \delta(F(\varphi))=1
%\ee usually the field $\varphi$ is a gauge-parameter, with \eqref{dirac_FP} determining the factor for a gauge-fixing. 
  
    Let a gauge-field $A^I_a(x)$, where $I$ are internal indices,  $a$ are spacetime ones, and we will suppress both,  on which a gauge-group $G$ acts as $A\rightarrow A^g$. For the gauge-fixing $C^\alpha(A)=0$, we define again a partition of unity by:
   \be\label{right_meas} 1=\Delta_C[A]\int d\mu_r(g)\delta C^\alpha(A^g)
   \ee
 where $d\mu_r$ is the right-invariant measure of the group $G$.   The previous equation can in fact be seen as a definition of      $\Delta_C[A]$, which implies it is invariant under a gauge transformation, $\Delta[C[A^g]]=\Delta[C[A]]$. 	
 Fadeev-Popov now tells us that an ensuing path integral decomposes into an integral over the gauge-group, and another, gauge-fixed invariant one, containing $\delta [C^\alpha(A)]\Delta_C[A]$. 
 
 Now,  if $g_o$ is such that $C[A^{g_o}]=0$, i.e., it satisfies the gauge-fixing condition, then following equation \eqref{delta_identity} we have that for an infinitesimal group translation, 
$$\delta_\epsilon C_{|g_o}=M_C\cdot \epsilon $$
where $M_C=\diby{C[A^{g_o+\epsilon}]}{\epsilon}$. Now, separating the total derivative along a vertical (i.e. along the gauge-direction) and horizontal (along the gauge-section, infinitesimally), $\delta=\delta_v+\delta_h$ (see \cite{Aldo_HG} and appendix \ref{app:PFB}) since then $\delta_h C[A]=0$, we have: 
 \be\label{left_meas} \int d\mu_\ell(g)\delta C^\alpha(A^g)=\det {M_C}(A)
   \ee
 
 From \eqref{left_meas} and \eqref{right_meas}, if the two measures coincide, we of course obtain $\Delta_C=\det {M_C}^{-1}$, as before, and thus $\det {M_C}$ is gauge-invariant . But this is not necessarily the case. Let $\tau^a$ be the canonical generators of the Lie algebra $\mathfrak{g}$. Then for any $g$, we have $\Lambda_a(g)$ such that  $g=\exp{(\Lambda_a\tau^a)}$. Then the difference between the left and the right measure is given by 
 $$d\mu_r(g)=d\mu_\ell(g) \exp{\varpi(g)}$$
 where $\varpi(g)=\Lambda_b(g)U^{ba}_a$. 
 Then we obtain from \eqref{right_meas}
 $$ 1=\Delta_C[A]\int d\mu_\ell(g) \exp{\varpi(g)}\delta C^\alpha(A^g)
 $$
and thus the correction to the relation between $\det {M_C}^{-1}$ and $\Delta_C$ becomes:
$$\delta_\epsilon \det {M_C}^{-1}=(1+\delta_\epsilon \Lambda_b(g)U^{ba}_a)\det {M_C}^{-1}
$$
 which only vanishes when the structure constants are field independent and traceless. This is our case with the algebra of $\mathcal{C}$, but it is not the case with the ADM algebra. 																																																																																																																																			  

%\bibliographystyle{alpha}
%\bibliography{decoherence}

\begin{thebibliography}{BBHV{\etalchar{+}}16}

\bibitem[Abb17]{GW_NS}
B.~P. et~al Abbott.
\newblock Gw170817: Observation of gravitational waves from a binary neutron
  star inspiral.
\newblock {\em Phys. Rev. Lett.}, 119:161101, Oct 2017.

\bibitem[ADM62]{ADM}
R.~Arnowitt, S.~Deser, and C.~Misner.
\newblock The dynamics of general relativity pp.227�264,.
\newblock In {\em in Gravitation: an introduction to current research, L.
  Witten, ed.} Wiley, New York, 1962.

\bibitem[Bar94]{Barbour94}
Julian Barbour.
\newblock The timelessness of quantum gravity: I the evidence from the
  classical theory.
\newblock {\em Class. Quant. Grav.}, 11:2853--2873, 1994.

\bibitem[BBHV{\etalchar{+}}16]{Barvinsky_hor}
Andrei~O. Barvinsky, Diego Blas, Mario Herrero-Valea, Sergey~M. Sibiryakov, and
  Christian~F. Steinwachs.
\newblock {Renormalization of Horava gravity}.
\newblock {\em Phys. Rev.}, D93(6):064022, 2016.

\bibitem[Ber96]{Bertlmann}
R.A. Bertlmann.
\newblock {\em Anomalies in quantum field theory}.
\newblock Cambridge University Press,, 1996.

\bibitem[BHM13]{Michor_general}
Martin Bauer, Philipp Harms, and Peter~W. Michor.
\newblock Sobolev metrics on the manifold of all riemannian metrics.
\newblock {\em J. Differential Geom.}, 94(2):187--208, 06 2013.

\bibitem[BKM14]{Barbour_arrow}
Julian Barbour, Tim Koslowski, and Flavio Mercati.
\newblock {Identification of a gravitational arrow of time}.
\newblock {\em Phys. Rev. Lett.}, 113(18):181101, 2014.

\bibitem[Ble81]{Bleecker}
David Bleecker.
\newblock {\em Gauge Theory and Variational Principles}.
\newblock Dover Publications, 1981.

\bibitem[CLMZ06]{BSSN_binary}
Manuela Campanelli, C.~O. Lousto, P.~Marronetti, and Y.~Zlochower.
\newblock {Accurate evolutions of orbiting black-hole binaries without
  excision}.
\newblock {\em Phys. Rev. Lett.}, 96:111101, 2006.

\bibitem[DeW67]{DeWitt_QG1}
Bryce~S. DeWitt.
\newblock {Quantum Theory of Gravity. 1. The Canonical Theory}.
\newblock {\em Phys. Rev.}, 160:1113--1148, 1967.

\bibitem[DF16]{Donnelly_2016}
William Donnelly and Laurent Freidel.
\newblock {Local subsystems in gauge theory and gravity}.
\newblock {\em JHEP}, 09:102, 2016.

\bibitem[DG17]{Donnelly_local}
William Donnelly and Steven~B. Giddings.
\newblock {How is quantum information localized in gravity?}
\newblock {\em Phys. Rev.}, D96(8):086013, 2017.

\bibitem[Ebi70]{Ebin}
D.G Ebin.
\newblock The manifold of riemmanian metrics.
\newblock {\em Symp. Pure Math., AMS,}, 11,15, 1970.

\bibitem[Fac13]{Faci_CG}
Sofiane Faci.
\newblock {Constructing conformally invariant equations by using Weyl
  geometry}.
\newblock {\em Class. Quant. Grav.}, 30:115005, 2013.

\bibitem[FG89]{Freed}
Daniel~S. Freed and David Groisser.
\newblock {The basic geometry of the manifold of Riemannian metrics and of its
  quotient by the diffeomorphism group.}
\newblock {\em Mich. Math. J.}, 36(3):323--344, 1989.

\bibitem[Fis70]{Fischer}
Arthur Fischer.
\newblock The theory of superspace.
\newblock In {\em Proceedings of the Relativity Conference held 2-6 June, 1969
  in Cincinnati, OH. Edited by Moshe Carmeli, Stuart I. Fickler, and Louis
  Witten. New York: Plenum Press, 1970., p.303}, 1970.

\bibitem[FM77]{FiMa77}
A.~Fischer and J.~Marsden.
\newblock The manifold of conformally equivalent metrics.
\newblock {\em Can. J. Math.}, 29:193--209, (1977).

\bibitem[FM02]{Fischer_Marsden}
Arthur~E. Fischer and Vincent Moncrief.
\newblock {\em Conformal Volume Collapse of 3 Manifolds and the Reduced
  Einstein Flow}, pages 463 -- 522.
\newblock Springer New York, New York, NY, 2002.

\bibitem[GGK11]{SD_first}
Henrique Gomes, Sean Gryb, and Tim Koslowski.
\newblock {Einstein gravity as a 3D conformally invariant theory}.
\newblock {\em Class. Quant. Grav.}, 28:045005, 2011.

\bibitem[GHHM04]{Cotton_squashed}
Alberto Garcia, Friedrich~W. Hehl, Christian Heinicke, and Alfredo Macias.
\newblock {The Cotton tensor in Riemannian space-times}.
\newblock {\em Class. Quant. Grav.}, 21:1099--1118, 2004.

\bibitem[Giu09]{Giulini_geometrodynamics}
Domenico Giulini.
\newblock {The Superspace of Geometrodynamics}.
\newblock {\em Gen. Rel. Grav.}, 41:785--815, 2009.

\bibitem[GK11]{SD_coupling}
Henrique Gomes and Tim Koslowski.
\newblock Coupling shape dynamics to matter gives spacetime.
\newblock {\em Gen. Rel. Grav. 44, No 6 (2012), 1539-1553}, 10 2011.

\bibitem[GK12]{SD_linking}
Henrique Gomes and Tim Koslowski.
\newblock {The Link between General Relativity and Shape Dynamics}.
\newblock {\em Class.Quant.Grav.}, 29:075009, 2012.

\bibitem[GKMN15]{Thin_shell}
Henrique Gomes, Tim Koslowski, Flavio Mercati, and Andrea Napoletano.
\newblock {Gravitational collapse of thin shells of dust in asymptotically flat
  Shape Dynamics}.
\newblock 2015.

\bibitem[GMM91]{Gil_Medrano}
Olga Gil-Medrano and Peter~W. Michor.
\newblock The riemannian manifold of all riemannian metrics.
\newblock {\em Quarterly Journal of Mathematics, (42), 183-202}, 1991.

\bibitem[Gom11]{Gauge_riem}
Henrique Gomes.
\newblock Gauge theory in riem: Classical.
\newblock {\em J. Math. Phys. 52, 082501}, 2011.

\bibitem[Gom13]{SD_Weyl}
Henrique Gomes.
\newblock {A first look at Weyl anomalies in shape dynamics}.
\newblock {\em J. Math. Phys.}, 54:112302, 2013.

\bibitem[Gom14]{SD_birkhoff}
Henrique Gomes.
\newblock {A Birkhoff theorem for Shape Dynamics}.
\newblock {\em Class.Quant.Grav.}, 31:085008, 2014.

\bibitem[Gom15]{SD_construction}
Henrique Gomes.
\newblock {Conformal geometrodynamics regained: gravity from duality}.
\newblock {\em Annals Phys. 355, 224-240}, 355:224--240, 2015.

\bibitem[Gom17]{QG_deco}
Henrique Gomes.
\newblock {Quantum gravity in timeless configuration space}.
\newblock {\em Class. Quant. Grav.}, 34(23):235004, 2017.

\bibitem[Gou07]{3+1_book}
Eric Gourgoulhon.
\newblock {3+1 formalism and bases of numerical relativity}.
\newblock {\em Lecture notes in Physics 846, Springer}, 2007.

\bibitem[GR17]{Aldo_HG}
Henrique Gomes and Aldo Riello.
\newblock The observer's ghost: notes on a field space connection.
\newblock {\em Journal of High Energy Physics}, 2017(5):17, May 2017.

\bibitem[GS16]{Vasu_rigidity}
Henrique Gomes and Vasudev Shyam.
\newblock {Extending the rigidity of general relativity}.
\newblock {\em J. Math. Phys.}, 57(11):112503, 2016.

\bibitem[HE73]{Hawking_Ellis}
S.~Hawking and G.~Ellis.
\newblock {\em The Large Scale Structure of Space-Time}.
\newblock Cambridge Univ. Press, 1973.

\bibitem[Her16]{Gabe_parity}
Gabriel Herczeg.
\newblock {Parity horizons in shape dynamics}.
\newblock {\em Class. Quant. Grav.}, 33(22):225002, 2016.

\bibitem[HH83]{Hartle_Hawking}
S.W. Hawking and J.~Hartle.
\newblock {Wave function of the Universe}.
\newblock {\em Phys. Rev. D, Vol 28, Issue 12}, 28:2960--2975, December 1983.

\bibitem[HH91]{Hartle_Halliwell}
Jonathan~J. Halliwell and James~B. Hartle.
\newblock Wave functions constructed from an invariant sum over histories
  satisfy constraints.
\newblock {\em Phys. Rev. D}, 43:1170--1194, Feb 1991.

\bibitem[Hor09]{Horava}
Petr Horava.
\newblock {Quantum Gravity at a Lifshitz Point}.
\newblock {\em Phys.Rev.}, D79:084008, 2009.

\bibitem[IM82]{IsMa1982}
James Isenberg and Jerrold Marsden.
\newblock A slice theorem for the space of solutions of einstein's equations.
\newblock {\em Phys. Rep., 89}, 1982.

\bibitem[IMJWY76]{York3}
J.~A. Isenberg, N.~Murchadha, and Jr. J.~W.~York.
\newblock Initial value problem of general relativity. 3. coupled fields and
  the scalar-tensor theory.
\newblock {\em Phys. Rev. D , 13, 1532-1537.}, 1976.

\bibitem[IN77]{IsenbergNester2}
J.~A. Isenberg and J.~M. Nester.
\newblock Extension of the york field decomposition to general gravitationally
  coupled fields,.
\newblock {\em Annals Phys. (108), 368-386}, 1977.

\bibitem[Jas87]{Jaskolski}
Z.~Jaskolski.
\newblock The integration of $g$-invariant functions and the geometry of the
  faddeev-popov procedure.
\newblock {\em Comm. Math. Phys.}, 111(3):439--468, 1987.

\bibitem[JG15]{Horndeski}
F.~Piazza F.~Vernizzi; J.~Gleyzes, D.~Langlois.
\newblock Healthy theories beyond horndeski.
\newblock {\em Phys. Rev. Lett. 114}, 2015.

\bibitem[JWY73]{York2}
Jr. J.~W.~York.
\newblock Conformally invariant orthogonal decomposition of symmetric tensors
  on riemannian manifolds and the initial value problem of general relativity,.
\newblock {\em J.\ Math.\ Phys. {\bf{14}} 456.}, 1973.

\bibitem[KMS18]{Through_BB}
Tim~A. Koslowski, Flavio Mercati, and David Sloan.
\newblock Through the big bang: Continuing einstein's equations beyond a
  cosmological singularity.
\newblock {\em Physics Letters B}, 778:339 -- 343, 2018.

\bibitem[Lin84]{Linde}
A~D Linde.
\newblock The inflationary universe.
\newblock {\em Reports on Progress in Physics}, 47(8):925, 1984.

\bibitem[LM97]{Lavelle_review}
Martin Lavelle and David McMullan.
\newblock {Constituent quarks from QCD}.
\newblock {\em Phys. Rept.}, 279:1--65, 1997.

\bibitem[LP87]{Yamabe}
John~M. Lee and Thomas~H. Parker.
\newblock The yamabe problem.
\newblock {\em Bull. Amer. Math. Soc.}, 17:37--91, 1987.

\bibitem[LW90]{Wald_Lee}
Joohan Lee and Robert~M. Wald.
\newblock Local symmetries and constraints.
\newblock {\em Journal of Mathematical Physics}, 31(3):725--743, 1990.

\bibitem[Max14]{Maxwell}
David Maxwell.
\newblock {The Conformal Method and the Conformal Thin-Sandwich Method Are the
  Same}.
\newblock {\em Class. Quant. Grav.}, 31:145006, 2014.

\bibitem[Mer16]{Flavio_single}
Flavio Mercati.
\newblock On the fate of birkhoff's theorem in shape dynamics.
\newblock {\em accepted for publication by Gen. Rel. Grav.}, 2016.

\bibitem[Mer17a]{Flavio_tutorial}
Flavio Mercati.
\newblock Shape dynamics: Relativity and relationalism.
\newblock {\em Oxford University Press}, 2017.

\bibitem[Mer17b]{Flavio_compact}
Flavio Mercati.
\newblock {Thin shells of dust in a compact universe}.
\newblock 2017.

\bibitem[Mur89]{Yamabe_Murchadha}
Niall~. Murchadha.
\newblock The yamabe theorem and general relativity.
\newblock In {\em Conference on Mathematical Relativity}, pages 137--167,
  Canberra AUS, 1989. Centre for Mathematics and its Applications, Mathematical
  Sciences Institute, The Australian National University.

\bibitem[OJ73]{Niall_York}
Niall O'Murchadha and James W.~York Jr.
\newblock Existence and uniqueness of solutions of the hamiltonian constraint
  of general relativity on compact manifolds.
\newblock {\em Journal of Mathematical Physics}, 14(11):1551--1557, 1973.

\bibitem[Pal61]{Palais}
R.~Palais.
\newblock On the existence of slices for the actions of non-compact groups.
\newblock {\em Ann. of Math.}, 73:295--322, 1961.

\bibitem[Pan08]{Paneitz}
Stephen Paneitz.
\newblock A quartic conformally covariant differential operator for arbitrary
  pseudo-riemannian manifolds (summary).
\newblock {\em SIGMA 4 036}, (2008).

\bibitem[Pre05]{Pretorius_binary}
Frans Pretorius.
\newblock {Evolution of binary black hole spacetimes}.
\newblock {\em Phys. Rev. Lett.}, 95:121101, 2005.

\bibitem[RS13]{Path_foliations}
Stefan Rechenberger and Frank Saueressig.
\newblock {A functional renormalization group equation for foliated
  spacetimes}.
\newblock {\em JHEP}, 03:010, 2013.

\bibitem[Sin78]{Singer_Gribov}
I.~M. Singer.
\newblock {Some remarks on the Gribov ambiguity}.
\newblock {\em Communications in Mathematical Physics}, 60(1):7--12, feb 1978.

\bibitem[Tei83]{Teitelboim_path}
Claudio Teitelboim.
\newblock Proper-time gauge in the quantum theory of gravitation.
\newblock {\em Phys. Rev. D}, 28:297--309, Jul 1983.

\bibitem[Vil86]{Vilenkin}
Alexander Vilenkin.
\newblock Boundary conditions in quantum cosmology.
\newblock {\em Phys. Rev. D}, 33:3560--3569, Jun 1986.

\bibitem[Wey18]{Weyl}
H.~Weyl.
\newblock �gravitation und elektrizit�t�.
\newblock {\em pp. 147-59 dans Das Relativit�tsprinzip}, 1918.

\bibitem[Wil95]{Wiltshire_intro}
David~L. Wiltshire.
\newblock {An Introduction to quantum cosmology}.
\newblock In {\em {Cosmology: The Physics of the Universe. Proceedings, 8th
  Physics Summer School, Canberra, Australia, Jan 16-Feb 3, 1995}}, pages
  473--531, 1995.

\bibitem[Yor71]{York}
James~W. York.
\newblock Gravitational degrees of freedom and the initial-value problem.
\newblock {\em Phys. Rev. Lett.}, 26:1656--1658, 1971.

\end{thebibliography}

\newcommand{\etalchar}[1]{$^{#1}$}

\end{document}